\documentclass[11pt,eqs]{article}
\usepackage{latexsym}

\def\cleq{\setcounter{equation}{0}}
\begin{document}
\title{
Non-commutativity parameters depend
not only on the effective coordinate
but on its T-dual as well
\thanks{Work supported in part by the Serbian Ministry of Science and
Technological Development, under contract No. 171031.}}
\author{Lj. Davidovi\'c \thanks{e-mail address: ljubica@ipb.ac.rs} and B. Sazdovi\'c
\thanks{e-mail address: sazdovic@ipb.ac.rs}\\
{\it Institute of Physics,}\\
{\it University of Belgrade,}\\
{\it 11001 Belgrade, P.O.Box 57, Serbia}}
\maketitle
\begin{abstract}
We extend our investigations of the open string propagation
in the weakly curved background to the case when Kalb-Ramond field,
beside the infinitesimal term linear in coordinate $B_{\mu\nu\rho}x^\rho$,
contains the constant term $b_{\mu\nu}\neq 0$.
In two previously investigated cases,
for the flat background ($b_{\mu\nu}\neq 0$ and $B_{\mu\nu\rho}=0$) and the weakly curved one
($b_{\mu\nu}= 0$ and $B_{\mu\nu\rho}\neq 0$)
the effective metric is constant and the effective Kalb-Ramond field is zero.
In the present article ($b_{\mu\nu}\neq 0$ and $B_{\mu\nu\rho}\neq 0$)
the effective metric is coordinate dependent and there exists
non-trivial effective Kalb-Ramond field.
It depends on the $\sigma$-integral
of the effective momenta $P_\mu(\sigma)=\int_{0}^{\sigma}d\eta p_\mu(\eta)$,
which is in fact T-dual of the effective coordinate, $P_\mu=\kappa g_{\mu\nu}{\tilde{q}}^\nu$.
Beside the standard coordinate dependent term
$\theta^{\mu\nu}(q)$,
in the non-commutativity parameter,
which is nontrivial only on the string end-points,
there are additional $P_\mu$ (or ${\tilde{q}}^\mu$)
dependent terms which are nontrivial
both at the string endpoints and at the string interior.
The additional terms are infinitesimally small.
The part of one of these terms
has been obtained in Ref. \cite{HY} and the others are our improvements.
\end{abstract}
%%%%%%%%%%%%%%%%%%%%%%%%%%%%%%%%%%%%%%%%%%%%%%%%%%%%%%%%%%%%%%%%%%%%%%%%%%%%%%%%%%%%%%%%%%%%%%%%%%%%%%%%%%%%%%%%%%%%%%%%%%%%%%%%%%%%%%%%%%%
\section{Introduction}

It is well known that in the presence of
the Kalb-Ramond antisymmetric tensor field $B_{\mu\nu}$,
quantization of the open string ending on Dp-branes
leads to the non-commutativity of the string end-points
\cite{CDS}-\cite{DS}.

In the majority of papers the case of a flat background,
with constant metric tensor $G_{\mu\nu}$,
antisymmetric tensor $B_{\mu\nu}$ and the dilaton field $\Phi$
has been investigated.
The constant $B_{\mu\nu}$ field
is a source of non-commutativity
at the string end-points and it does
not affect dynamics in the world sheet interior.
Several methods have been used to investigate this case:
operator product expansion of the open string vertex
operator \cite{Sch,VS}, the mode expansion of the
classical solution \cite{AC}, the methods of conformal field theory
\cite{SW} and the canonical quantization
for constrained systems \cite{ACL,SN}.

The case of the linear dilaton field,
is similar to that of the constant background
and was considered in Refs. \cite{SN}.
Non-constant dilaton field induces a commutative Dp-brane coordinate
in the direction of dilaton gradient $\partial_\mu \Phi$.
When $\partial_\mu \Phi$ is lightlike vector
with respect to the open or to the closed string metric,
the local gauge symmetries appear. They turn some Neumann boundary conditions
into Dirichlet ones and decrease the number of Dp-brane dimensions \cite{SN}.

In Refs. \cite{JdB} the Dp-brane embedded in  IIB superstring theory
space-time was considered. The presence of the $\Omega$-odd fields,
antisymmetric under world-sheet parity transformation ($\Omega:\sigma\rightarrow -\sigma$)
leads to non-commutativity of the supercoordinates.
This non-commutative supermultiplet beside $B_{\mu\nu}$ from NS-NS sector
contains difference of two gravitons $\psi^{\alpha}_{-\mu}$
from NS-R sector and symmetric part of bispinor $F^{\alpha\beta}$ from R-R sector.

In the great majority of investigations, the target space was assumed to be flat.
Working with curved target space is
extremely complicated and in all papers some assumptions were made.
In Refs. \cite{VS,CSS,HKK,ARS,DS} the solution of the space-time equation,
known as {\it weakly curved} background
has been used.
Weakly curved background means that the metric tensor $G_{\mu\nu}$ is constant,
the antisymmetric tensor $B_{\mu\nu}$ is linear in coordinate and
its field strength $B_{\mu\nu\rho}$ is infinitesimally small non-vanishing parameter.
From the spacetime equation of motion follows that
the Ricci curvature $R_{\mu\nu}$
is an infinitesimal of the second order,
and as such is neglected.

In order to investigate the open string with non-vanishing field strength of Kalb-Ramond field,
in Refs. \cite{CSS,HKK}
the correlation functions have been computed on the disc and therefrom the Kontsevich product has been extracted.
Considerations in Ref. \cite{HKK}
has been restricted to the first order in derivatives of the background fields,
while in the Ref. \cite{CSS} it has been
restricted to the weakly curved background.

In Refs. \cite{HKK,ARS} the non-commutativity parameter depends
on the coordinate and has the standard form.
In the Ref. \cite{HY}
in the low energy limit,
the new kind of non-commutativity
relation has been obtained, where
non-commutativity parameter depends not only on the coordinate,
but on the momentum as well.
This form of parameter has not been
observed by the path integral method, in Refs. \cite{CSS,HKK}.
To resolve the discrepancies of these results,
in the present paper
we will extend the systematic canonical
approach of the Ref. \cite{DS}, to the case
$b_{\mu\nu}\equiv B_{\mu\nu}[x=0]\neq 0$.
We show that the momentum dependent terms
of the Ref. \cite{HY} really exist
and that they disappear for $b_{\mu\nu}= 0$.
We also find new momentum
dependent terms, missed in \cite{HY} as a consequence of low
energy limit assumptions.

In the present paper we investigate the propagation
of the open string in the weakly curved background using canonical methods.
The approach of our previous paper \cite{DS},
which was applied to the
curved background is generalized
to the case when the constant part of the
Kalb-Ramond field is not equal to zero, $b_{\mu\nu}\neq 0$.
Treating the boundary conditions as constraints and
using Dirac requirement that the time derivatives of the constraints are also constraints,
we obtain the infinite set of constraints in the
Lagrangian form.
Following the line of Refs. \cite{SN,DS},
we represent this infinite set of constraints at string end-point
with $\sigma-$dependent constraints.
As well as in Ref. \cite{DS}
we succeed to express these constraints in the compact form.

Switching from the Lagrangian to the Hamiltonian method,
we show that the Poisson bracket between Hamiltonian and the constraints
are again the constraints.
Therefore, they are in fact Hamiltonian constraints and they
form a complete set of constraints.

All constraints except the zero modes \cite{BS} are of the second class and we solve them explicitly.
They appear as particular orbifold conditions reducing the initial phase space to the $\Omega$-even
and $2\pi-$periodic one.
Instead of using the Dirac brackets in the initial phase space associated with
the second class constraints,
we will use the equivalent star brackets in the effective
phase space \cite{DS}.

The initial coordinate $x^\mu$ depends
both on the effective coordinate $q^\mu$ and its canonically
conjugated momentum $p_\mu$. As well as in \cite{DS},
this fact is a source of non-commutativity.
But unlike \cite{DS} where the coefficient in front of $p_\mu$
was infinitesimally small,
now this coefficient contains
constant term $\theta^{\mu\nu}_{0}$,
which is not infinitesimal.
As a consequence, the non-commutativity parameter
depends not only on $q^\mu$ as in \cite{DS},
but on both $q^\mu$ and $p_\mu$.

Let us clarify notation and terminology used in the
two descriptions of the open string theory.
All parameters of the initial description
with variables $x^\mu$ and $\pi_\mu$
and background fields $G_{\mu\nu}$ and $B_{\mu\nu}$,
following Seiberg and Witten \cite{SW},
we will call {\it closed string parameters}.
Such a theory is given by the equations of motion and
the boundary conditions.
On the solution of the boundary conditions,
we obtain the string theory
defined on the $2\pi$-periodic, $\Omega$-symmetric subspace and
given by the equations of motion only.
The corresponding variables $q^\mu$ and $p_\mu$
and background fields
$G_{\mu\nu}^{eff}[q]$, $B^{eff}_{\mu\nu}[P]$
will be called the {\it effective parameters}.
In analogy with Seiberg and Witten,
we will refer to the mappings
$G^{E}_{\mu\nu}(G,B)$ and $\theta^{\mu\nu}(G,B)$
 defined in (\ref{eq:gefv}) and (\ref{eq:teta})
as the {\it open string background fields}.
Note that in our case, there are two sets
of the open string backgrounds,
one in terms of the closed string background fields
$G^{E}_{\mu\nu}(G,B)$ and $\theta^{\mu\nu}(G,B)$,
and the other in terms of the effective background fields
$G^{E}_{\mu\nu}(G^{eff},B^{eff})$ and
$\theta^{\mu\nu}(G^{eff},B^{eff})$
(see appendix \ref{sec:osbf}).

%%%%%%%%%%%%%%%%%%%%%%%%%%%%%%%%%%%%%%%%%%%%%%%%%%%%%%%%%%%%%%%%%%%%%%%%%%%%%%%%%%%%%%%%%%%%%%%%%%%%%%%%%%%%%%%%%%%
\section{Definition of the model}

We will investigate the open bosonic string, in the nontrivial background
defined by the space-time fields:
the metric $G_{\mu\nu}$ and the Kalb-Ramond antisymmetric tensor $B_{\mu\nu}$.
The propagation is
described by the action \cite{FTFCB,GSW,ZW}
\begin{equation}\label{eq:action0}
S = \kappa \int_{\Sigma} d^2\xi\sqrt{-g}
\Big[\frac{1}{2}{g}^{\alpha\beta}G_{\mu\nu}[x]
+\frac{\epsilon^{\alpha\beta}}{\sqrt{-g}}B_{\mu\nu}[x]\Big]
\partial_{\alpha}x^{\mu}\partial_{\beta}x^{\nu},
\quad (\varepsilon^{01}=-1),
\end{equation}
where integration goes over two-dimensional world-sheet $\Sigma$
with coordinates $\xi^{0}=\tau,\ \xi^{1}=\sigma$ and
$g_{\alpha\beta}$ is intrinsic world-sheet metric.
By $x^{\mu}(\xi),\ \mu=0,1,...,D-1$ we denote the coordinates of the D-dimensional space-time,
and we use the notation
$\dot{x}=\frac{\partial x}{\partial\tau}$,
$x^\prime=\frac{\partial x}{\partial\sigma}$.
Choosing the conformal gauge $g_{\alpha\beta}=e^{2F}\eta_{\alpha\beta}$
the action takes the form
\begin{equation}\label{eq:action}
S = \kappa \int_{\Sigma} d^2\xi
\Big[\frac{1}{2}\eta^{\alpha\beta}G_{\mu\nu}[x]
+{\epsilon^{\alpha\beta}}B_{\mu\nu}[x]\Big]
\partial_{\alpha}x^{\mu}\partial_{\beta}x^{\nu}.
\end{equation}

In the case of the open string, the minimal action principle produces
the equation of motion
\begin{equation}\label{eq:motion}
{\ddot{x}}^\mu=x^{\prime\prime\mu}
-2B^\mu_{\ \nu\rho}{\dot{x}}^\nu x^{\prime\rho},
\end{equation}
and the boundary conditions on the string endpoints
\begin{equation}\label{eq:bonc}
\gamma^{0}_\mu\Big{|}_{\sigma=0,\pi}=0,
\end{equation}
where we have introduced the variable
\begin{equation}
\gamma^{0}_{\mu}\equiv \frac{\delta {\cal{L}}}{\delta x^{\prime\mu}}
=G_{\mu\nu}x^{\prime\nu}-2B_{\mu\nu}\dot{x}^\nu.
\end{equation}

The consistency of the theory requires the world-sheet conformal invariance on the quantum level.
This means that the background fields satisfy the space-time equations of motion.
To the lowest order in slope parameter $\alpha^\prime$, for the constant dilaton field $\Phi=const$
these equations have the form
\begin{equation}\label{eq:beta}
R_{\mu \nu} - \frac{1}{4} B_{\mu \rho \sigma}
B_{\nu}^{\ \rho \sigma}=0\, ,
\end{equation}
\begin{equation}\label{eq:beta1}
D_\rho B^{\rho}_{\ \mu \nu} = 0,
\end{equation}
where
$B_{\mu\nu\rho}=\partial_\mu B_{\nu\rho}
+\partial_\nu B_{\rho\mu}+\partial_\rho B_{\mu\nu}$
is a field strength of the field $B_{\mu \nu}$, and
$R_{\mu \nu}$ and $D_\mu$ are Ricci tensor and
covariant derivative with respect to the space-time metric.

We will consider the following particular solution \cite{VS,CSS,DS,DS2}
\begin{eqnarray}\label{eq:gb}
G_{\mu\nu}=const,\quad
B_{\mu\nu}[x]=b_{\mu\nu}+\frac{1}{3}B_{\mu\nu\rho}x^\rho,
\end{eqnarray}
where the parameter $b_{\mu\nu}$ is constant and $B_{\mu\nu\rho}$
is constant and infinitesimally small.
Through the paper we will work up to the first order in $B_{\mu\nu\rho}$.
Note that the Ricci tensor $R_{\mu\nu}$ in (\ref{eq:beta})
is an infinitesimal of the second order,
and as such is neglected.
This background is known as {\it weakly curved} background.
Note that there is
an additional constant term $b_{\mu\nu}\neq 0$
in comparison with the solution of the Ref.\cite{DS},
which will produce a significant difference in the final result.
%%%%%%%%%%%%%%%%%%%%%%%%%%%%%%%%%%%%%%%%%%%%%%%%%%%%%%%%%%%%%%%%%%%%%%%%%%%%%%%%%%%%%%%%%%
\section{Compact form of the constraints}
\cleq
We will treat the boundary conditions (\ref{eq:bonc}), obtained
from the minimal action principle, as constraints. By demanding their
time consistency, we will obtain the infinite set of constraints.
At both string end-points these constraints will be substituted
with one $\sigma$-dependent constraint,
which will be rewritten explicitly in the canonical form.
At the end of the section we will
show that this constraints are of the second class.
The content of this section
is a generalization of the procedure of the Ref. \cite{DS}
to the case $b_{\mu\nu}\neq 0$.
%%%%%%%%%%%%%%%%%%%%%%%%%%%%%%%%%%%%%%%%%%%%%%%%%%%%%%%%%%%%%%%%%%%%%%%%%%%%%%%%%%%%%%%%%%
\subsection{Lagrangian consistency conditions}
\cleq

In order to obtain the explicit form of the infinite set of constraints
\begin{eqnarray}\label{eq:gaman}
\gamma^{n}_{\mu}\equiv{\dot{\gamma}}^{n-1}_{\mu},\quad({n\geq 1})
\end{eqnarray}
it is useful to introduce the following variables
\begin{eqnarray}\label{eq:functions}
\gamma_\mu=G_{\mu\nu}x^{\prime\nu}-2B_{\mu\nu}\dot{x}^\nu,&&
{\tilde{\gamma}}_\mu=G_{\mu\nu}\dot{x}^\nu-2B_{\mu\nu}x'^\nu,
\nonumber\\
Q^{n}_{\mu}=B_{\mu\nu\rho}{\dot{x}}^{(n)\nu}x^{(n+1)\rho},&&
R^{n}_{\mu}=B_{\mu\nu\rho}\Big{[}
x^{(n+2)\nu}x^{(n+1)\rho}
+{\dot{x}}^{(n)\nu}{\dot{x}}^{(n+1)\rho}\Big{]},
\end{eqnarray}
where $x^{(n)\mu}\equiv\frac{{\partial}^{n}}{\partial{\sigma}^{n}}x^\mu$.
On the equation of motion (\ref{eq:motion}),
in the leading order
the first time derivatives of these functions
have a form
\begin{eqnarray}
{\dot{\gamma}}_\mu={\tilde{\gamma}}^{\prime}_{\mu}
+4b_{\mu}^{\ \nu}{Q}^{0}_{\nu},
&&
{\dot{\tilde{\gamma}}}_\mu={\gamma}^{\prime}_{\mu}
-\frac{2}{3}Q^{0}_{\mu},
\nonumber\\
{\dot{Q}}^{n}_{\mu}=R^{n}_{\mu},
&&
{\dot{R}}^{n}_{\mu}=(Q^{n}_{\mu})^{\prime\prime}
-4Q^{n+1}_{\mu},
\end{eqnarray}
and the second time derivatives are closed on the same set of variables
\begin{eqnarray}
{\ddot{\gamma}}_\mu={\gamma}^{\prime\prime}_{\mu}
-\frac{2}{3}(Q^{0}_{\mu})^{\prime}
+4b_{\mu}^{\ \nu}{R}^{0}_{\nu},
&&
{\ddot{\tilde{\gamma}}}_\mu=
{\tilde{\gamma}}^{\prime\prime}_{\mu}
-\frac{2}{3}R^{0}_{\mu}+4b_{\mu}^{\ \nu}
(Q_{\nu}^{0})^{\prime},
\nonumber\\
{\ddot{Q}}^{n}_{\mu}=
(Q^{n}_{\mu})^{\prime\prime}
-4Q^{n+1}_{\mu},
&&
{\ddot{R}}^{n}_{\mu}=
(R^{n}_{\mu})^{\prime\prime}
-4R^{n+1}_{\mu}.
\end{eqnarray}

Following the procedure developed in our paper \cite{DS}, we obtain
that the constraints are equal to
\begin{eqnarray}
\gamma^{0}_{\mu}=\gamma_{\mu},
&&
{\gamma}^{1}_\mu={\tilde{\gamma}}^{\prime}_{\mu}
+4b_{\mu}^{\ \nu}{Q}^{0}_{\nu}.
\end{eqnarray}
\begin{eqnarray}
\gamma^{2n}_{\mu}&=&\gamma^{(2n)}_{\mu}
-\frac{2}{3}\sum_{k=0}^{n-1}\alpha^{k}_{2n}
(Q^{k}_{\mu})^{(2n-2k-1)}
+4b_{\mu}^{\ \nu}
\sum_{k=0}^{n-1}\alpha^{k}_{2n}
(R^{k}_{\nu})^{(2n-2k-2)},
\quad (n\geq{1})
\nonumber\\
\gamma^{2n+1}_{\mu}&=&{\tilde{\gamma}}^{(2n+1)}_{\mu}
-\frac{2}{3}
\sum_{k=0}^{n-1}\alpha^{k}_{2n}
(R^{k}_{\mu})^{(2n-2k-1)}
+4b_{\mu}^{\ \nu}
\sum_{k=0}^{n}\alpha^{k}_{2n+2}
(Q^{k}_{\nu})^{(2n-2k)},\quad (n\geq{1})
\nonumber\\
\end{eqnarray}
where
\begin{equation}
\alpha^{k}_{2n}=(-4)^{k}{n \choose {k+1}},\qquad {k=0,\cdots,n-1}\ .
\end{equation}
Instead of working with the infinite number of constraints,
we form
two $\sigma$-dependent constraints
\begin{eqnarray}\label{eq:gamasa}
\Gamma^{S}_{\mu}(\sigma)\equiv\sum_{n=0}^{\infty}\frac{\sigma^{2n}}{(2n)!}
\gamma^{2n}_{\mu}\Big{|}_{\sigma=0}=0,&&
\Gamma^{A}_{\mu}(\sigma)\equiv\sum_{n=0}^{\infty}\frac{\sigma^{2n+1}}{(2n+1)!}
\gamma^{2n+1}_{\mu}\Big{|}_{\sigma=0}=0.
\end{eqnarray}
They have the form
\begin{eqnarray}\label{eq:gsa}
\Gamma^{S}_{\mu}(\sigma)
&=&\gamma^{S}_{\mu}(\sigma)
-\frac{2}{3}
\Gamma^{Q}_{\mu}(\sigma)
+4b_{\mu}^{\ \nu}
{\tilde{\Gamma}}^{R}_{\nu}(\sigma),
\nonumber\\
\Gamma^{A}_{\mu}(\sigma)
&=&{\tilde{\gamma}}^{A}_{\mu}(\sigma)
-\frac{2}{3}\Gamma^{R}_{\mu}(\sigma)
+4b_{\mu}^{\ \nu}{\tilde{\Gamma}}^{Q}_{\nu}(\sigma),
\end{eqnarray}
with $\gamma^{S}_{\mu}(\sigma)\equiv
{\sum_{n=0}^{\infty}\frac{\sigma^{2n}}{(2n)!}
\gamma^{(2n)}_{\mu}}\Big{|}_{\sigma=0},\quad$
${\tilde{\gamma}}^{A}_{\mu}(\sigma)\equiv
{\sum_{n=0}^{\infty}\frac{\sigma^{2n+1}}{(2n+1)!}
\gamma^{(2n+1)}_{\mu}}\Big{|}_{\sigma=0},\quad$
and where we have defined
\begin{eqnarray}
\Gamma^{Q}_{\mu}
&\equiv&
\sum_{n=1}^{\infty}\sum_{k=0}^{n-1}\frac{\sigma^{2n}}{(2n)!}
\alpha^{k}_{2n}(Q^{k}_{\mu})^{(2n-2k-1)}\Big{|}_{\sigma=0}
\nonumber\\
&=&
\sum_{k=0}^{\infty}
\sum_{n=k+1}^{\infty}\frac{\sigma^{2n}}{(2n)!}
\alpha^{k}_{2n}(Q^{k}_{\mu})^{(2n-2k-1)}\Big{|}_{\sigma=0},
\\
\Gamma^{R}_{\mu}
&\equiv&
\sum_{n=1}^{\infty}\sum_{k=0}^{n-1}\frac{\sigma^{2n+1}}{(2n+1)!}
\alpha^{k}_{2n}(R^{k}_{\mu})^{(2n-2k-1)}\Big{|}_{\sigma=0}
\nonumber\\
&=&
\sum_{k=0}^{\infty}\sum_{n=k+1}^{\infty}\frac{\sigma^{2n+1}}{(2n+1)!}
\alpha^{k}_{2n}(R^{k}_{\mu})^{(2n-2k-1)}\Big{|}_{\sigma=0},
\\
{\tilde{\Gamma}}^{R}_{\mu}&\equiv&
\sum_{n=1}^{\infty}\sum_{k=0}^{n-1}\frac{\sigma^{2n}}{(2n)!}
\alpha^{k}_{2n}(R^{k}_{\mu})^{(2n-2k-2)}\Big{|}_{\sigma=0}
\nonumber\\
&=&\sum_{k=0}^{\infty}\sum_{n=k+1}^{\infty}\frac{\sigma^{2n}}{(2n)!}
\alpha^{k}_{2n}(R^{k}_{\mu})^{(2n-2k-2)}\Big{|}_{\sigma=0},
\\
{\tilde{\Gamma}}^{Q}_{\mu}
&\equiv&
\sum_{n=0}^{\infty}\sum_{k=0}^{n}\frac{\sigma^{2n+1}}{(2n+1)!}
\alpha^{k}_{2n+2}
(Q^{k}_{\mu})^{(2n-2k)}\Big{|}_{\sigma=0}
\nonumber\\
&=&\sum_{k=0}^{\infty}\sum_{n=k}^{\infty}\frac{\sigma^{2n+1}}{(2n+1)!}
\alpha^{k}_{2n+2}(Q^{k}_{\mu})^{(2n-2k)}\Big{|}_{\sigma=0}.
\end{eqnarray}
The definitions (\ref{eq:gaman}) and (\ref{eq:gamasa})
produce the following relation
\begin{eqnarray}
{\dot{\Gamma}}^{S}_{\mu}=\Gamma^{\prime A}_{\mu},
\end{eqnarray}
which is in terms of the components equal to
\begin{eqnarray}
{\dot{\gamma}}^{S}_{\mu}&=&{\tilde{\gamma}}^{\prime A}_{\mu}+
4b_{\mu}^{\ \nu}(Q^{0}_{\nu})^{S},
\nonumber\\
{\dot{\Gamma}}^{Q}_{\mu}&=&
{\Gamma}^{\prime R}_{\mu},
\nonumber\\
({\tilde{\Gamma}}^{R}_{\mu})^{\cdot}&=&
{\tilde{\Gamma}}^{\prime Q}_{\mu}
-(Q^{0}_{\mu})^{S}.
\end{eqnarray}
By index S(A) we mark the parts of the function, with even (odd)
powers in sigma.
The quantities $\Gamma^{Q},\Gamma^{R},{\tilde{\Gamma}}^{R},{\tilde{\Gamma}}^{Q}$
can be rewritten in the integral form
(for details see appendix C of Ref.\cite{DS})
\begin{equation}\label{eq:gqint}
\Gamma^{Q}_{\mu}(\sigma)=\frac{\sigma}{2}
\sum_{k=0}^{\infty}
\frac{(-1)^{k}}{(k+1)!}
\int_{0}^{\sigma}d\sigma_{1}^{2}
\int_{0}^{\sigma_{1}}d\sigma_{2}^{2}\cdots
\int_{0}^{\sigma_{k-1}}d\sigma_{k}^{2}
(Q^{k}_{\mu})^{A}
(\sigma_{k}),
\end{equation}
\begin{eqnarray}\label{eq:grint}
\Gamma^{R}_{\mu}(\sigma)&=&
\frac{1}{4}
\sum_{k=0}^{\infty}
\frac{(-1)^{k}}{(k+1)!}
\int_{0}^{\sigma}d\sigma_{0}^{2}
\int_{0}^{\sigma_{0}}d\sigma_{1}^{2}\cdots
\int_{0}^{\sigma_{k-1}}d\sigma_{k}^{2}
(R^{k}_{\mu})^{A}(\sigma_{k}),
\end{eqnarray}
\begin{equation}\label{eq:grtint}
{\tilde{\Gamma}}^{R}_{\mu}(\sigma)=
\frac{\sigma}{2}\sum_{k=0}^{\infty}
\frac{(-1)^{k}}{(k+1)!}
\int_{0}^{\sigma}d\sigma_{1}^{2}
\int_{0}^{\sigma_{1}}d\sigma_{2}^{2}\cdots
\int_{0}^{\sigma_{k-1}}d\sigma_{k}^{2}
\int_{0}^{\sigma_{k}}d\eta
(R^{k}_{\mu})^{S}(\eta),
\end{equation}
\begin{equation}\label{eq:gqtint}
{\tilde{\Gamma}}^{Q}_{\mu}(\sigma)=
\frac{\partial}{\partial\sigma}
\Big{[}\frac{\sigma}{2}\sum_{k=0}^{\infty}
\frac{(-1)^{k}}{(k+1)!}
\int_{0}^{\sigma}d\sigma_{1}^{2}
\int_{0}^{\sigma_{1}}d\sigma_{2}^{2}\cdots
\int_{0}^{\sigma_{k-1}}d\sigma_{k}^{2}
\int_{0}^{\sigma_{k}}d\eta
(Q^{k}_{\mu})^{S}(\eta)\Big{]}.
\end{equation}

Separating coordinates at
even and odd parts $x^\mu=q^\mu+{\bar{q}}^\mu$
\begin{eqnarray}\label{eq:qqbar}
q^\mu(\sigma)=
\sum_{n=0}^{\infty}\frac{{\sigma}^{2n}}{(2n)!}x^{(2n)\mu}\Big{|}_{\sigma=0},&&
{\bar{q}}^\mu(\sigma)=
\sum_{n=0}^{\infty}\frac{{\sigma}^{2n+1}}{(2n+1)!}x^{(2n+1)\mu}\Big{|}_{\sigma=0},
\end{eqnarray}
with the help of $Z$ summation formula derived in appendix \ref{sec:zsum}
we can rewrite (\ref{eq:grtint}) and (\ref{eq:gqtint})
in the compact form
\begin{eqnarray}
{\tilde{\Gamma}}^{R}_{\mu}(\sigma)&=&
B_{\mu\nu\rho}
\Big{[}
i^{\nu\rho}[q^{\prime\prime},{\bar{q}}]
+i^{\nu\rho}[{\bar{q}}^{\prime\prime},q]
+i^{\nu\rho}[\dot{q},\dot{\bar{q}}]
+i^{\nu\rho}[\dot{\bar{q}},\dot{q}]
\Big{]}
\nonumber\\
&=&B_{\mu\nu\rho}
\Big{[}
\frac{1}{2}q^{\prime\nu}{\bar{q}}^\rho
+\frac{1}{2}{\dot{Q}}^\nu{\dot{\bar{q}}}^{\rho}
\Big{]},
\nonumber\\
{\tilde{\Gamma}}^{Q}_{\mu}(\sigma)&=&
B_{\mu\nu\rho}
\frac{\partial}{\partial\sigma}\Big{[}i^{\nu\rho}[\dot{q},\bar{q}]
+i^{\nu\rho}[\dot{\bar{q}},q]\Big{]}
=B_{\mu\nu\rho}
\frac{\partial}{\partial\sigma}
\Big{[}{\dot{Q}}^\nu {\bar{q}}^\rho\Big{]},
\end{eqnarray}
where  the functions $i^{\nu\rho}$ are defined in (\ref{eq:idef}).
In the paper \cite{DS} we obtained
\begin{eqnarray}\label{eq:sqr}
\Gamma^{Q}_{\mu}(\sigma)&=&
B_{\mu\nu\rho}\Big{[}
h^{\nu\rho}[\dot{q},q]+h^{\nu\rho}[\dot{\bar{q}},{\bar{q}}]
\Big{]}
=\frac{1}{2}B_{\mu\nu\rho}\Big{[}
\dot{Q}^\nu q ^{\prime\rho}+\frac{1}{2}{\dot{\bar{q}}}^\nu{\bar{q}}^\rho
\Big{]},
\nonumber\\
\Gamma^{R}_{\mu}(\sigma)&=&
B_{\mu\nu\rho}
\int_{0}^{\sigma}d\sigma_{0}
\Big{[}h^{\nu\rho}[q^{\prime\prime},q]+h^{\nu\rho}[{\bar{q}}^{\prime\prime},{\bar{q}}]
+
h^{\nu\rho}[\dot{q},\dot{q}]+h^{\nu\rho}[\dot{\bar{q}},\dot{\bar{q}}]\Big{]}(\sigma_{0})
\nonumber\\
&=&\frac{1}{2}B_{\mu\nu\rho}
\int_{0}^{\sigma}d\sigma_{0}
\Big{[}
{\bar{q}}^{\prime\prime\nu}{\bar{q}}^\rho
+{\dot{Q}}^\nu{\dot{q}}^{\prime\rho}
\Big{]}.
\end{eqnarray}

Therefore,
the compact form of the constraints is equal to
\begin{eqnarray}\label{eq:const}
\Gamma_{\mu}^{S}(\sigma)&=&
G_{\mu\nu}{\bar{q}}^{\prime\nu}
-2b_{\mu\nu}{\dot{q}}^\nu
-\frac{2}{3}B_{\mu\nu\rho}
\Big{[}{\dot{q}}^\nu q^\rho
+\frac{1}{2}{\dot{Q}}^\nu q^{\prime\rho}
+\frac{3}{2}{\dot{\bar{q}}}^\nu {\bar{q}}^\rho\Big{]}
+2b_{\mu}^{\ \nu}B_{\nu\rho\upsilon}
\Big{[}q^{\prime\rho} {\bar{q}}^\upsilon
+{\dot{Q}}^\rho {\dot{\bar{q}}}^\upsilon\Big{]},
\nonumber\\
\Gamma_{\mu}^{A}(\sigma)&=&
G_{\mu\nu}{\dot{\bar{q}}}^\nu
-2b_{\mu\nu}{q}^{\prime\nu}
-\frac{2}{3}B_{\mu\nu\rho}
\Big{[}q^{\prime\nu}q^\rho
+\frac{1}{2}{\dot{Q}}^\nu{\dot{q}}^\rho
+\frac{3}{2}{\bar{q}}^{\prime\nu}{\bar{q}}^\rho\Big{]}
+2b_{\mu}^{\ \nu}B_{\nu\rho\upsilon}
\frac{\partial}{\partial\sigma}
\Big{[}{\dot{Q}}^\rho {\bar{q}}^\upsilon\Big{]},
\nonumber\\
\end{eqnarray}
where
\begin{equation}
 {Q}^\mu(\sigma)=\int_{0}^{\sigma}d\eta q^\mu(\eta).
\end{equation}

%%%%%%%%%%%%%%%%%%%%%%%%%%%%%%%%%%%%%%%%%%%%%%%%%%%%
\subsection{Canonical form of the constraints}
Ones we obtained the compact form of the constraints,
we can rewrite them in the canonical form.
The momenta corresponding to the coordinates $x^\mu$
have the form
\begin{equation}\label{eq:pi}
\pi_\mu=\kappa(G_{\mu\nu}{\dot{x}}^\nu-2B_{\mu\nu}x^{\prime\nu}).
\end{equation}
Extracting the time derivative of the
even and odd parts of the coordinate,
and introducing
\begin{equation}
p_\mu(\sigma)=\sum_{k=0}^{\infty}\frac{\sigma^{2k}}{(2k)!}
\pi_\mu^{(2k)}\Big{|}_{\sigma=0},\quad
{\bar{p}}_\mu(\sigma)=\sum_{k=0}^{\infty}\frac{\sigma^{2k+1}}{(2k+1)!}
\pi_\mu^{(2k+1)}\Big{|}_{\sigma=0},
\end{equation}
we find
\begin{eqnarray}
{\dot{q}}^\mu&=&\frac{1}{\kappa}(G^{-1})^{\mu\nu}p_\nu
+2b^{\mu}_{\ \nu}{\bar{q}}^{\prime\nu}
+\frac{2}{3}B^{\mu}_{\ \nu\rho}({\bar{q}}^{\prime\nu}{q}^\rho+q^{\prime\nu}{\bar{q}}^\rho),
\nonumber\\
{\dot{\bar{q}}}^\mu&=&\frac{1}{\kappa}(G^{-1})^{\mu\nu}{\bar{p}}_\nu
+2b^{\mu}_{\ \nu}{q}^{\prime\nu}
+\frac{2}{3}B^{\mu}_{\ \nu\rho}({\bar{q}}^{\prime\nu}{\bar{q}}^\rho+q^{\prime\nu}{q}^\rho).
\end{eqnarray}
Substituting these relations
into the expression (\ref{eq:const}),
we obtain
the constraints in the canonical form
\begin{eqnarray}\label{eq:gs}
\Gamma_{\mu}^{S}(\sigma)&=&
G^{E}_{\mu\nu}[q]{\bar{q}}^{\prime\nu}
-\frac{2}{\kappa}B_{\mu}^{\ \nu}[q]p_\nu
\nonumber\\
&+&\Big{[}
-2(bh+hb)+6h[bq]+24bh[bq]b\Big{]}^{\prime}_{\mu\nu}{\bar{q}}^\nu
\nonumber\\
&-&\frac{1}{\kappa}\Big{[}h-12bh[bq]\Big{]}_{\mu}^{\prime\ \nu}P_\nu
\nonumber\\
&-&\frac{3}{\kappa}\Big{[}\bar{h}+4b{h}[b\bar{q}]\Big{]}_{\mu}^{\ \nu}{\bar{p}}_\nu
+\frac{6}{{\kappa}^{2}}\Big{[}b{h}[\bar{p}]\Big{]}_{\mu}^{\ \nu}P_\nu,
\end{eqnarray}
\begin{eqnarray}\label{eq:ga}
\Gamma_{\mu}^{A}(\sigma)&=&
\frac{1}{\kappa}{\bar{p}}_\mu
\nonumber\\
&+&\Big{[}-\bar{h}+12b\bar{h}b+4h[b\bar{q}]b-12bh[b\bar{q}]
\Big{]}_{\mu\nu}{\bar{q}}^{\prime\nu}
\nonumber\\
&+&\frac{2}{\kappa}\Big{[}3b\bar{h}+h[b\bar{q}]\Big{]}_{\mu}^{\ \nu}p_\nu
\nonumber\\
&+&\frac{2}{\kappa}\Big{[}
3b\bar{h}-h[b\bar{q}]\Big{]}_{\mu}^{\prime \nu}P_\nu
-\frac{1}{\kappa^{2}}h_{\mu}^{\ \nu}[p]P_\nu,
\end{eqnarray}
where $G^{E}_{\mu\nu}$ and $h$,$\bar{h}$ are defined
in (\ref{eq:gefv}) and (\ref{eq:hhbar}), and
\begin{equation}
P_\mu(\sigma)=\int_{0}^{\sigma}d\eta p_\mu(\eta).
\end{equation}

%%%%%%%%%%%%%%%%%%%%%%%%%%%%%%%%%%%%%%%%%%%%%%%%%%%%%%%%%%%%%%%%%%%%%%%%%%%%%%%%%%%%%%%%%%%%%%%
\subsection{Second class constraints}
In order to find to which class the constraints belong,
first we find the canonical Hamiltonian
${H}_{c}=\int_{0}^{\pi}d\sigma{\cal{H}}_{c}(x,\pi)$,
corresponding to the action (\ref{eq:action}).
The Hamiltonian density equals
\begin{equation}\label{eq:hamc}
{\cal{H}}_{c}(x,\pi)=\frac{1}{2\kappa}\pi_\mu (G^{-1})^{\mu\nu}\pi_\nu
+\frac{\kappa}{2}x^{\prime\mu}G^{E}_{\mu\nu}[x]x^{\prime\nu}
-2x^{\prime\mu}B_{\mu\nu}[x](G^{-1})^{\nu\rho}\pi_\rho.
\end{equation}
Note that from the standard Poisson brackets
\begin{equation}
\{x^\mu(\sigma),\pi_\nu(\bar{\sigma})\}=\delta^{\mu}_{\nu}\delta(\sigma-\bar{\sigma}),
\end{equation}
we have two non-trivial relations for $\Omega$ even and odd subspaces
\begin{eqnarray}\label{eq:poisson}
\{q^\mu(\sigma),p_\nu(\bar{\sigma})\}=2
\delta^{\mu}_{\nu}\delta_{S}(\sigma,\bar{\sigma}),
\quad
\{{\bar{q}}^\mu(\sigma),{\bar{p}}_\nu(\bar{\sigma})\}=2
\delta^{\mu}_{\nu}\delta_{A}(\sigma,\bar{\sigma}),
\end{eqnarray}
where $\delta_{S}$ and $\delta_{A}$ are defined in (\ref{eq:dsa}).
The reasons for appearance
of the factor 2 can be found in the App. B
of the Ref. \cite{DS}.

Because the
symmetric and the antisymmetric
constraints $\Gamma_{\mu}^{S}$ and $\Gamma_{\mu}^{A}$
are independent,
it is equivalent to consider the constraint
$\Gamma_{\mu}=\kappa(\Gamma_{\mu}^{S}+\Gamma_{\mu}^{A})$.
It weakly commutes with the Hamiltonian
\begin{equation}
\{ H_c, \Gamma^\mu (\sigma) \} = \Gamma^{\prime\mu} (\sigma),
\end{equation}
and therefore, there are no more constraints.

Next,
we will calculate the Poisson bracket
$\{\Gamma_{\mu}(\sigma),\Gamma_{\nu}(\bar{\sigma})\}$,
considering only
the part which depends on the
derivatives of the delta functions.
It equals
\begin{eqnarray}\label{eq:poisgama}
\{\Gamma_{\mu}(\sigma),\Gamma_{\nu}(\bar{\sigma})\}&=&
\kappa\Big{[}
G^{E}_{\mu\nu}[q(\bar{\sigma})]
+\kappa X_{\mu\nu}[\bar{q}(\sigma)]+\kappa Y_{\nu\mu}[\bar{q}(\bar{\sigma})]
\Big{]}
\delta_{S}^{\prime}(\sigma,\bar{\sigma})
\nonumber\\
&+&\kappa\Big{[}
G^{E}_{\mu\nu}[q({\sigma})]
+\kappa X_{\nu\mu}[\bar{q}(\bar{\sigma})]+\kappa Y_{\mu\nu}[\bar{q}({\sigma})]
\Big{]}
\delta_{A}^{\prime}(\sigma,\bar{\sigma})+\cdots
\nonumber\\
&=&\kappa\Big{[}
G^{E}_{\mu\nu}[q(\sigma)]
+\kappa(X+Y)_{\mu\nu}[\bar{q}(\sigma)]
+\kappa(X+Y)_{\nu\mu}[\bar{q}(\sigma)]
\Big{]}\delta^{\prime}(\sigma-\bar{\sigma})+\cdots,
\nonumber\\
\end{eqnarray}
where we defined
\begin{eqnarray}
X_{\mu\nu}&=&
-\frac{3}{\kappa}{\bar{h}}_{\mu\nu}
+\frac{4}{\kappa}(b\bar{h}b)_{\mu\nu}
+Z_{\mu\nu},
\nonumber\\
Y_{\mu\nu}&=&
-\frac{1}{\kappa}{\bar{h}}_{\mu\nu}
+\frac{12}{\kappa}(b\bar{h}b)_{\mu\nu}
+Z_{\mu\nu},
\nonumber\\
Z_{\mu\nu}&=&
\frac{1}{\kappa}\Big{[}
4h[b\bar{q}]b-12bh[b\bar{q}]
+\frac{2}{\kappa}h[P]b
\Big{]}_{\mu\nu}.
\end{eqnarray}
After some calculation it reduces to
\begin{eqnarray}
\{\Gamma_{\mu}(\sigma),\Gamma_{\nu}(\bar{\sigma})\}&=&
\kappa\Big{[}
G^{E}_{\mu\nu}[q(\sigma)]+Z_{\mu\nu}[\bar{q}(\sigma)]+Z_{\nu\mu}[\bar{q}(\sigma)]
\Big{]}\delta^{\prime}(\sigma-\bar{\sigma})+\cdots
\nonumber\\
&=&\kappa G_{\mu\nu}^{E}[(q+2b\bar{q}+\frac{1}{\kappa}P)(\sigma)]
\delta^{\prime}(\sigma-\bar{\sigma})+\cdots.
\end{eqnarray}
Supposing that the open string
metric $G^{E}_{\mu\nu}$ is regular,
we can conclude that
all the constraints $\Gamma_{\mu}$
except the zero modes are the
second class constraints.

%%%%%%%%%%%%%%%%%%%%%%%%%%%%%%%%%%%%%%%%%%%%%%%%%%
\section{Solution of the constraints}
\cleq

Similarly as in Ref. \cite{DS} we should also
solve the constraints at $\sigma=\pi$.
Because the orientation of the string is
the matter of choice,
it is clear that the constraints at $\sigma=\pi$
are related to that at $\sigma=0$.
As well as in Refs.\cite{SN,DS} the $2\pi-$periodicity
of the original variables
$x^\mu(\sigma+2\pi)=x^\mu(\sigma)$, $\pi_\mu(\sigma+2\pi)=\pi_\mu(\sigma)$
and the solution of the constraints at $\sigma=0$
are enough to solve the constraints at $\sigma=\pi$.

We solve the constraints at $\sigma=0$ by the iteration method.
In the zeroth order in  the small parameter $B_{\mu\nu\rho}$,
the solutions of the symmetric
and of the antisymmetric constraints are
\begin{eqnarray}\label{eq:qbar}
{\bar{q}}_{0}^{\prime\mu}=-\theta^{\mu\nu}_{0}p_\nu,
\quad\rightarrow\quad
{\bar{q}}_{0}^{\mu}=-\theta^{\mu\nu}_{0}P_\nu,
\end{eqnarray}
and
\begin{equation}
{\bar{p}}^{0}_\mu=0,
\end{equation}
where $\theta^{\mu\nu}_{0}$ is defined in (\ref{eq:gtetao}).

Substituting the above relations into the
infinitesimal part of the constraint (\ref{eq:gs}),
with the help of (\ref{eq:pomoc}) we obtain
\begin{equation}\label{eq:gs1}
\Gamma_{\mu}^{S}(\sigma)=
G^{E}_{\mu\nu}[q]
\Big{[}
{\bar{q}}^{\prime\nu}
+\theta^{\nu\rho}[q]p_\rho
+\frac{1}{2}{{\Lambda}}_{-}^{\prime\nu\rho}[q]P_\rho
\Big{]},
\end{equation}
where
$G^{E}_{\mu\nu}$ , $\theta^{\mu\nu}$ and
${{\Lambda}}^{\mu\nu}_{-}$
are defined in (\ref{eq:gefv}) and (\ref{eq:teta}) and (\ref{eq:lambda}).
Similarly we find
\begin{eqnarray}\label{eq:ga1}
\Gamma_{\mu}^{A}(\sigma)&=&
\frac{1}{\kappa}{\bar{p}}_\mu
+\Big{[}
-h[\theta_{0}P]\theta_{0}
-\frac{6}{\kappa}bh[\theta_{0}P]g^{-1}
\nonumber\\
&+&\frac{1}{\kappa^{2}}h[g^{-1}P]g^{-1}
+\frac{6}{\kappa}bh[g^{-1}P]\theta_{0}
\Big{]}_{\mu}^{\ \nu}p_\nu,
\end{eqnarray}
where $g_{\mu\nu}$ and $\theta_{0}^{\mu\nu}$ are introduced in (\ref{eq:gtetao}).

So, on the solution of these constraints
$\Gamma^{S}_{\mu}(\sigma)=0$
and $\Gamma^{A}_{\mu}(\sigma)=0$,
up to the terms linear in $B_{\mu\nu\rho}$ we have
\begin{eqnarray}\label{eq:xpres}
x^\mu(\sigma)&=&q^\mu(\sigma)+{\bar{q}}^{\mu}(\sigma)
\nonumber\\
&=&q^\mu(\sigma)-\int_{0}^{\sigma}d\eta\Big{[}
\theta^{\mu\nu}[q(\eta)]p_\nu(\eta)
+\frac{1}{2}{{\Lambda}}^{\prime\mu\nu}_{-}[q(\eta)]P_\nu(\eta)\Big{]},
\nonumber\\
\pi_\mu(\sigma)&=&p_\mu(\sigma)+{\bar{p}}_\mu(\sigma)
\nonumber\\
&=&p_\mu+
\Big{[}Gb^{-1}\beta[{\bar{q}}_{0}]g^{-1}\Big{]}_{\mu}^{\ \nu}p_\nu.
\end{eqnarray}
In the last equality we introduced the new variable
$\beta_{\mu\nu}[\bar{q_{0}}]$
which is infinitesimally small as well as
${\bar{p}}_\mu$.
Explicitly, it is equal to
\begin{equation}\label{eq:betadef}
\beta_{\mu\nu}[{\bar{q}}_{0}]=
2\Big{[}
bh[{\bar{q}}_{0}]b-3b^{2}h[{\bar{q}}_{0}]
-\frac{1}{4}b{{h}}[b^{-1}{\bar{q}}_{0}]
+3b^{2}{h}[b^{-1}{\bar{q}}_{0}]b
\Big{]}_{\mu\nu}.
\end{equation}

It is essential that the term $\theta^{\mu\nu}[q]$,
in the expression (\ref{eq:xpres}) for ${\bar{q}}^\mu$,
in addition to that of Ref. \cite{DS},
contains the finite part $\theta^{\mu\nu}_{0}$.
Therefore, ${\bar{q}}^\mu$ is not
infinitesimal, and
the canonical brackets $^{\star}\{{\bar{q}}^\mu,{\bar{q}}^\nu\}$
will produce nontrivial momenta dependent improvement.

%%%%%%%%%%%%%%%%%%%%%%%%%%%%%%%%%%%%%%%%%%%%%%
\section{The effective theory}
\cleq
In the previous section we obtained the solution (\ref{eq:xpres}) of
the boundary conditions (\ref{eq:gs}) and (\ref{eq:ga}),
with the initial
canonical variables $x^\mu$ and $\pi_\mu$
given in terms of the effective ones $q^\mu$ and $p_\mu$.
In this section we will
constraint the initial theory with
this solution,
in order to obtain the effective one.
From the requirement that the initial and the effective
Hamiltonian and Lagrangian should
have the same form, we can find the effective background fields.
We also show that the effective
Kalb-Ramond field
does not depend on the effective coordinate
but on its T-dual.
 %%%%%%%%%%%%%%%%%%%%%%%%%%%%%%%%%%%%%%%%%%%%%%%%%%%%%%%%%
\subsection{The Effective Hamiltonian}\label{sec:effham}

The effective theory is described by the effective Lagrangian
\begin{equation}\label{eq:leff}
{\cal{L}}^{eff}(q,p)=[\pi_\mu{\dot{x}}^\mu-{\cal{H}}_{c}(x,\pi)]
\Big{|}_{\Gamma_{\mu}=0}.
\end{equation}
Because the basic canonical variables $q^\mu(\sigma)$
and $p_\mu(\sigma)$
contain only the even powers of $\sigma$,
we will extend their domain to $\sigma\in[-\pi,\pi]$ and
consider $q^\mu$ and $p_\mu$ as even functions on that interval.
Consequently,
we will consider the action
\begin{equation}
S^{eff}=\int d\tau \int_{-\pi}^{\pi}d\sigma{\cal{L}}^{eff},
\end{equation}
which means that we perform the $\Omega$-even projection
of the ${\cal{L}}^{eff}$
and consider the unoriented effective theory.

By substituting the solution (\ref{eq:xpres}) into the symmetrized first
term of the  expression (\ref{eq:leff}) and into (\ref{eq:hamc}) we obtain
\begin{equation}\label{eq:betadis}
\frac{1}{2}(I+
\Omega)
\Big{[}\pi_\mu{\dot{x}}^\mu
\Big{]}\Big{|}_{\Gamma_{\mu}=0}
=p_\mu{\dot{q}}^\mu-2q^{\prime\mu}
\Big{[}
\beta[{\bar{q}}_{0}]g^{-1}
\Big{]}_\mu^{\ \nu}p_\nu,
\end{equation}
and
\begin{eqnarray}\label{eq:hamcon}
{{\cal{H}}_{c}}(x,\pi)\Big{|}_{
\Gamma_{\mu}=0}
&=&\frac{1}{2\kappa}p_\mu (G^{-1}_{E})^{\mu\nu}[q]\;p_\nu
+\frac{\kappa}{2}q^{\prime\mu}G^{E}_{\mu\nu}[q]\;q^{\prime\nu}
\nonumber\\
&-&2q^{\prime\mu}
\Big{[}
\beta[{\bar{q}}_{0}]g^{-1}
+(\bar{h}+4b\bar{h}b)g^{-1}
\Big{]}_{\mu}^{\ \nu} p_{\nu}.
\end{eqnarray}

Note that the second term of (\ref{eq:betadis}) cancels
the corresponding one in the third term of (\ref{eq:hamcon}),
so we get the standard form of the action
\begin{equation}\label{eq:acteff}
S^{eff}=\int_{\Sigma_{1}}d\xi
\Big{[}
p_\mu{\dot{q}}^\mu
-{{\cal{H}}}^{eff}_{c}(q,p)
\Big{]},
\end{equation}
where $\Sigma_{1}$ means the integration over $\sigma$
in the interval $[-\pi,\pi]$ and
\begin{equation}\label{eq:hamc1}
{{\cal{H}}}^{eff}_{c}(q,p)=\frac{1}{2\kappa}p_\mu
(G^{-1}_{E})^{\mu\nu}[q]\;p_\nu
+\frac{\kappa}{2}q^{\prime\mu}
G^{E}_{\mu\nu}[q]\;q^{\prime\nu}
-2q^{\prime\mu}
\Big{[}(\bar{h}+4b\bar{h}b)g^{-1}\Big{]}_{\mu}^{\ \nu} p_{\nu}.
\end{equation}

The effective theory (and consequently the effective
Hamiltonian ${{\cal{H}}}^{eff}_{c}(q,p)$),
should depend on the effective variables $q^\mu,p_\mu$
in exactly the same way as the original theory (the original
Hamiltonian ${{\cal{H}}}_{c}(x,\pi)$) depends on original variables $x^\mu,\pi_\mu$.
From the first and the third term of (\ref{eq:hamc}) and
(\ref{eq:hamc1}) we can find the effective background fields
\begin{eqnarray}\label{eq:gbtoeff}
G_{\mu\nu}&\rightarrow&G^{E}_{\mu\nu}[q]\equiv{{G}}^{eff}_{\mu\nu}[q]
\nonumber\\
B_{\mu\nu}[x]&\rightarrow&
(\bar{h}+4b\bar{h}b)_{\mu\nu}
=-\frac{\kappa}{2}(g\Delta\theta[\bar{q}_{0}]g)_{\mu\nu}
\equiv {{B}}^{eff}_{\mu\nu}[\bar{q}_{0}],
\end{eqnarray}
with $G^{E}_{\mu\nu}$ and ${\Delta\theta}^{\mu\nu}$ defined
in (\ref{eq:gefv}) and (\ref{eq:deltag}).

So, the effective background fields are in fact the open
string background fields (\ref{eq:gefv}) and (\ref{eq:teta}).
The antisymmetric one
contains only the infinitesimal part,
with the argument (\ref{eq:qbar}),
proportional to the
integral of the momenta.
Because it is infinitesimally small, the
transition between
the second terms of (\ref{eq:hamc}) and (\ref{eq:hamc1}),
is in agreement with (\ref{eq:gbtoeff}).
\begin{eqnarray}
{G}^{E}_{\mu\nu}(x)=G^{E}_{\mu\nu}(G,B)
\rightarrow
G^{E}_{\mu\nu}(G^{eff},B^{eff})
={{G}}^{eff}_{\mu\nu}[q]
=G^{E}_{\mu\nu}[q].
\end{eqnarray}

The complete transition from the original to the effective theory consists of\\
1. the {\it canonical variable} transition
\begin{equation}\label{eq:vt}
x^\mu,\pi_\mu\rightarrow q^\mu,p_\mu,
\end{equation}
2. the {\it background field} transition
\begin{eqnarray}\label{eq:gbefff}
G_{\mu\nu}\rightarrow G_{\mu\nu}^{eff}[q],
&&
B_{\mu\nu}[x]\rightarrow B^{eff}_{\mu\nu}[-\theta_{0}P].
\end{eqnarray}
Note that the effective background fields become
coordinate and momentum dependent.

For the flat initial background
($G_{\mu\nu}$ and $B_{\mu\nu}$ constant),
the effective background has constant effective metric
$g_{\mu\nu}$ and zero Kalb-Ramond field.
Infinitesimal correction of the initial antisymmetric field $B_{\mu\nu}$,
linear in coordinate $x^\mu$, produces the infinitesimal correction
to the effective metric, linear in the effective coordinate
$q^\mu$ and infinitesimal correction of the Kalb-Ramond field,
linear in $\sigma$ integral of the effective momenta
$P_\mu(\sigma)=\int_{0}^{\sigma}d\eta p_\mu(\eta)$.

It is well known that in the theory of the unoriented closed string
(which is just our effective theory) the Kalb-Ramond field vanishes.
Let us present the standard arguments
which support this fact \cite{SN,ZW}.
Because of the presence of $q^\prime$,
the term with Kalb-Ramond field
${{q}}^{\prime\mu} (Bg^{-1})_{\mu}^{\ \nu}p_\nu$,
changes the sign under $\Omega$-transformation
and consequently its integral over $\sigma$
in the symmetric interval $[-\pi,\pi]$ disappears.

Now, let us explain what is different in our case and how does the nontrivial
effective Kalb-Ramond background field $B^{eff}_{\mu\nu}[-\theta_{0}P]$
appear \cite{DS2}?
The difference is in the
fact that the effective Kalb-Ramond field does not depend
on the effective coordinate $q^\mu$ ($\Omega$-even),
but on the integral of the effective momenta
$P_\mu(\sigma)=\int_{0}^{\sigma}d\eta p_\mu(\eta)$
($\Omega$-odd).
As $B_{\mu\nu}^{eff}(-\theta_{0}P)$ is linear in $P_\mu$,
the effective Kalb-Ramond field
is odd under $\sigma$-parity transformation
$\Omega B_{\mu\nu}^{eff}[-\theta_{0}P(\sigma)]=
-B_{\mu\nu}^{eff}[-\theta_{0}P(\sigma)]$.
This makes the term with the
effective Kalb-Ramond field $\Omega$-even,
which allows its survival.

%%%%%%%%%%%%%%%%%%%%%%%%%%%%%%%%%%%%%%%%%%%%%%%%%%%%%%%%%%%%%%%%%%%%%%%%%%%%%%%%%%%%
\subsection{The Effective Lagrangian}

Using the equations of motion
with respect to the momenta,
we will eliminate them from the action (\ref{eq:acteff}),
in order to obtain the effective Lagrangian.
In our case this is not a straightforward calculation,
because background fields depend on momenta and therefore the equations
of motion are not linear. The fact that we
are working with the small parameter
simplifies the solution.

It is more appropriate to find the equation of motion with respect
to $P_\mu$, instead with respect to $p_\mu=P^{\prime}_\mu$.
It has a form
\begin{equation}\label{eq:mom}
p_\mu=\kappa g_{\mu\nu}{\dot{q}}^\nu+\Delta p_\mu,
\end{equation}
where $\Delta p_\mu$ is known infinitesimal correction,
but it does not contribute to the effective Lagrangian.
Substituting (\ref{eq:mom}) into (\ref{eq:acteff}) we obtain
\begin{equation}\label{eq:efl}
{\cal{L}}^{eff}(q)=\frac{\kappa}{2}{\dot{q}}^\mu
G_{\mu\nu}^{E}[q]{\dot{q}}^\nu
-\frac{\kappa}{2}q^{\prime\mu} G_{\mu\nu}^{E}[q]q^{\prime\nu}
+2\kappa q^{\prime\mu}{B}^{eff}_{\mu\nu}[2b\dot{Q}]{\dot{q}}^\nu,
\end{equation}
where $B^{eff}_{\mu\nu}$ is defined in (\ref{eq:gbtoeff}) and
we used
\begin{equation}
{\bar{q}}_{0}^\mu=-\theta_{0}^{\mu\nu}P_\nu
=2(G^{-1}b)^{\mu}_{\ \nu}\dot{Q}^{\nu},
\qquad Q^\mu(\sigma)\equiv\int_{0}^{\sigma}d\eta q^\mu(\eta).
\end{equation}

The result (\ref{eq:efl}) agrees with that of the Ref. \cite{DS2},
obtained completely in the Lagrangian approach.
The solution of Ref. \cite{DS2}
\begin{eqnarray}\label{eq:sollag}
{x}^{\prime\mu}&=&
{q}^{\prime\mu}
+2B^{\mu}_{\ \nu}[q]{\dot{q}}^\nu
-A^{\mu}_{\ \nu}[\dot{Q}]{q}^{\prime\nu},
\nonumber\\
{\dot{x}}^{\mu}&=&
{\dot{q}}^{\mu}
+2B^{\mu}_{\ \nu}[q]{q}^{\prime\nu}
- A^{\mu}_{\ \nu}[\dot{Q}]{\dot{q}}^{\nu},
\end{eqnarray}
with
\begin{eqnarray}
A_{\mu\nu}[\dot{Q}]&=&
h_{\mu\nu}[\dot{Q}]
-12b_{\mu}^{\ \rho}h_{\rho\sigma}[\dot{Q}]b^{\sigma}_{\ \nu}
-12h_{\mu\rho}[b\dot{Q}]b^{\rho}_{\ \nu}
+12b_{\mu}^{\ \rho}h_{\rho\nu}[b\dot{Q}],
\end{eqnarray}
is equivalent to the solution (\ref{eq:xpres}),
on the relation (\ref{eq:pi}).
Note that
$\beta_{\mu\nu}$ from (\ref{eq:xpres}) is related with
$A_{\mu\nu}$ as
\begin{equation}
\beta_{\mu\nu}[2b\dot{Q}]
+4\Big{[}bh[2b\dot{Q}]b
\Big{]}_{\mu\nu}
=-
\Big{[}bA(\dot{Q})\Big{]}_{\mu\nu}.
\end{equation}
In Ref. \cite{DS2},
the Lagrangian (\ref{eq:efl}) is obtained by
substituting (\ref{eq:sollag}) into (\ref{eq:action}),
and taking the $\Omega$-even projection.

With the similar consideration as in subsection
\ref{sec:effham},
we obtain the Lagrangian
transition expressions analog to (\ref{eq:vt})-(\ref{eq:gbefff})
\begin{eqnarray}\label{eq:xgb}
x^\mu\rightarrow{q^\mu},
\quad
G_{\mu\nu}\rightarrow G^{eff}_{\mu\nu}[q],
\quad
B_{\mu\nu}[x]\rightarrow B^{eff}_{\mu\nu}[2b\dot{Q}].
\end{eqnarray}
Note that in the Lagrangian approach the effective Kalb-Ramond
field does not depend on the  $\Omega$-even effective
coordinate  $q^\mu$ but on the $\Omega$-odd
integral over $\sigma$ of its
$\tau$-derivative,
${\dot{Q}}^\mu(\sigma)=
\int_{0}^{\sigma}d\eta{\dot{q}}^\mu(\eta)$.
Consequently, the theory we obtained is nonlocal.
%%%%%%%%%%%%%%%%%%%%%%%%%%%%%%%%%%%%%%%%%%%%%%%%%%%%%%%%%%%%%%%%%%%%%%%%%%%%%%%%%%%%%%%%%%%%%
\subsection{The effective Kalb-Ramond field in fact depends on the T-dual coordinates}
\label{sec:effkr}

Let us give the interpretation of the argument of the effective Kalb-Ramond field
$B^{eff}_{\mu\nu}[{\bar{q}}_{0}]$.
In the Hamiltonian approach, the argument is proportional to the integral of the effective momenta
${\bar{q}}^{\mu}_{0}=-{\theta}^{\mu\nu}_{0}P_\nu$
and in the Lagrangian one, it is proportional to the integral of the
$\tau$-derivative of the effective coordinate
${\bar{q}}^{\mu}_{0}=-2(G^{-1}b)^{\mu}_{\ \nu}{\dot{Q}}^{\nu}$.
We considered only the zeroth order value ${\bar{q}}^{\mu}_{0}$,
because ${\bar{q}}^{\mu}$ appears only as an argument
of the infinitesimally small field $B^{eff}_{\mu\nu}$.

It is well known
(see e.g. subsection 2.4.1 of \cite{GPR}),
that the canonical transformation
generated by the function
\begin{equation}
F(q,\tilde{q})=\kappa\int_{-\pi}^{\pi}
d\sigma q^{\prime\mu}g_{\mu\nu}{\tilde{q}}^\nu,
\end{equation}
interchanges the momenta and the $\sigma$-derivative
of the coordinate.
In our case we have
\begin{equation}
p_\mu=-\frac{\partial F}{\partial q^\mu}=
\kappa g_{\mu\nu}{\tilde{q}}^{\prime\nu},
\end{equation}
or
\begin{equation}\label{eq:pdual}
P_\mu=\kappa g_{\mu\nu}{\tilde{q}}^{\nu}.
\end{equation}
On the other hand, this transformation leads
to the T-dual theory, so that ${\tilde{q}}^{\mu}$
is T-dual effective coordinate and the argument of
$B^{eff}_{\mu\nu}$ is therefore
\begin{equation}
{\bar{q}}^{\mu}_{0}=2(G^{-1}b)^{\mu}_{\ \nu}{\tilde{q}}^{\nu}.
\end{equation}
More discussion on the T-duality as the canonical transformation
can be found at \cite{SJ}.

In the Lagrangian approach,
as was shown in Ref. \cite{DS2}, on the equations
of motion
$\dot{Q}^\mu(\sigma)$ is equal to the T-dual coordinate
$\tilde{q}^\mu(\sigma)$.
In fact, because $\dot{Q}$
appears only as an argument of the infinitesimally small
variable $B^{eff}_{\mu\nu}$,
it is enough to consider the zeroth order equation of motion
$\partial_{+}\partial_{-}q^\mu=0$,
with the solution
${q}^\mu(\sigma)=f^{\mu}(\sigma^{+})
+f^{\mu}(\sigma^{-})$,\
($\sigma^{\pm}=\tau\pm\sigma$).
From the relation
${\dot{q}}^\mu(\sigma)=f^{\prime\mu}(\sigma^{+})
-f^{\prime\mu}(\sigma^{-})$,\
we obtain
${\dot{Q}}^\mu(\sigma)=f^{\mu}(\sigma^{+})
-f^{\mu}(\sigma^{-})
=\tilde{q}^\mu(\sigma)$.

Therefore, the effective Hamiltonian
\begin{equation}
{\cal{H}}^{eff} =
\frac{1}{2\kappa}p_\mu (G^{-1}_{eff})^{\mu\nu}[q]p_\nu
+\frac{\kappa}{2}q^{\prime\mu}G^{eff}_{\mu\nu}[q]q^{\prime\nu}
-2q^{\prime\mu}B^{eff}_{\mu\rho}[2b\tilde{q}](G^{-1})^{\rho\nu}p_\nu,
\end{equation}
and the effective Lagrangian
\begin{equation}\label{eq:lageff}
{\cal{L}}^{eff} =\kappa\Big{[}
\frac{1}{2}{\eta}^{\alpha\beta}G_{\mu\nu}^{eff}[q]
+{\epsilon^{\alpha\beta}}B_{\mu\nu}^{eff}[2b\tilde{q}]\Big{]}
\partial_{\alpha}q^{\mu}\partial_{\beta}q^{\nu}
\end{equation}
describe the propagation of
the effective string (with $\Omega$-even variables)
in the effective background.
The effective metric tensor depends on the coordinate
${q}^\mu(\sigma)$,
while the effective Kalb-Ramond field depends on the
corresponding T-dual coordinate
$\tilde{q}^\mu(\sigma)$.

Note that the zeroth order of
both the Hamiltonian and the Lagrangian
solution, (\ref{eq:xpres}) and (\ref{eq:sollag}),
can be rewritten in terms of the T-dual
coordinate
\begin{equation}
x^\mu=q^\mu+2(G^{-1}b)^{\mu}_{\ \nu}{\tilde{q}}^\nu.
\end{equation}
The effective metric depends on the first term $q^\mu$
and the effective Kalb-Ramond field on the second term
$2b\tilde{q}$,
but we can formally rewrite both the effective Hamiltonian and the
effective action, as if the effective background fields
depend on the same argument
\begin{eqnarray}\label{eq:aceff}
H^{eff}_{c}&=&\int_{-\pi}^{\pi}d\sigma{\cal{H}}^{eff}_{c}
\nonumber\\
&=&\int_{-\pi}^{\pi}d\sigma
\Big{[}
\frac{1}{2\kappa}p_\mu (G^{-1}_{eff})^{\mu\nu}[x]p_\nu
+\frac{\kappa}{2}q^{\prime\mu}G^{eff}_{\mu\nu}[x]q^{\prime\nu}
-2q^{\prime\mu}B^{eff}_{\mu\rho}[x](G^{-1})^{\rho\nu}p_\nu
\Big{]},
\nonumber\\
S^{eff}&=&\kappa\int d\tau\int_{-\pi}^{\pi}d\sigma
\Big{[}
\frac{1}{2}{\eta}^{\alpha\beta}G_{\mu\nu}^{eff}[x]
+{\epsilon^{\alpha\beta}}B_{\mu\nu}^{eff}[x]\Big{]}
\partial_{\alpha}q^{\mu}\partial_{\beta}q^{\nu}.
\end{eqnarray}

The terms with $\Delta G^{eff}_{\mu\nu}[2b\tilde{q}]$ and $B^{eff}_{\mu\nu}[q]$
do not contribute, because they are the $\Omega$-odd terms integrated over the symmetric
interval $\sigma\in[-\pi,\pi]$.
The effective background fields are proportional to the open
string background fields (\ref{eq:gefv}) and (\ref{eq:teta}).
%%%%%%%%%%%%%%%%%%%%%%%%%%%%%%%%%%%%%%%%%%%%%%%%%%%%%%%%%%%%%%%%%%%%%%%%%%%%%%%%%%%%%%%%%%%%%
\section{Non-commutativity}
\cleq
In this section we will find the non-commutativity relation
of the space-time coordinates.
We will use the expression (\ref{eq:xpres})
for the closed string variable $x^\mu$
in terms of the effective string variables $q^\mu$ and $p_\mu$.
The effective variables are fundamental quantities
in the reduced phase space
obtained on the solution of the boundary conditions.
We will use the star brackets in the effective phase space \cite{DS},
with the basic relation
\begin{equation}
{}^\star\{q^\mu(\sigma),p_\nu(\bar{\sigma})\}
=2\delta^{\mu}_{\nu}\delta_{S}(\sigma,\bar{\sigma}),
\end{equation}
which are equivalent to the Dirac ones in the initial phase space.

The fact that in the solution (\ref{eq:xpres}) for the coordinate $x^\mu$,
$\theta^{\mu\nu}$
contains a finite part causes a nontrivial contribution of \ ${}^\star\{{\bar{q}}^\mu,{\bar{q}}^\nu\}$.
So, in the case $b_{\mu\nu}\neq 0$ the non-commutativity parameter obtains additional terms.
%%%%%%%%%%%%%%%%%%%%%%%%%%%%%%%%%%%%%%%%%%%%%%%%%%%%%%%%%%%%%%%%%%%%%%%%%%%%%%%%%%%%%%%%%%%%%%%
\subsection{Contributions from the term ${}^\star\{{\bar{q}}^\mu,{\bar{q}}^\nu\}$ }\label{sec:addterm}
Separating the
solution for the $\Omega$-odd space-time coordinate into two parts
${\bar{q}}^{\mu}={\bar{q}}_{1}^{\mu}+{\bar{q}}_{2}^{\mu}$,
where
\begin{eqnarray}
{\bar{q}}_{1}^{\mu}=-\int_{0}^{\sigma}d\eta\theta^{\mu\nu}[q(\eta)]p_\nu(\eta),
&&\quad
{\bar{q}}_{2}^{\mu}=-\frac{1}{2}\int_{0}^{\sigma}d\eta
{{\Lambda}}^{\prime\mu\nu}_{-}[q(\eta)]P_\nu(\eta),
\end{eqnarray}
we obtain
\begin{eqnarray}\label{eq:qbqb}
{}^\star\{{\bar{q}}_{1}^\mu(\sigma),{\bar{q}}_{1}^\nu(\bar{\sigma})\}&=&
-2\theta^{\mu\alpha}_{0}\partial_\alpha \theta^{\nu\rho}K_\rho(\bar{\sigma},\sigma)
+2\theta^{\nu\alpha}_{0}\partial_\alpha \theta^{\mu\rho}K_\rho(\sigma,\bar{\sigma}),
\nonumber\\
{}^\star\{{\bar{q}}_{1}^\mu(\sigma),{\bar{q}}_{2}^\nu(\bar{\sigma})\}&=&
-\theta^{\mu\alpha}_{0}\partial_{\alpha}{{\Lambda}}^{\nu\rho}_{-}J_\rho(\bar{\sigma},\sigma),
\end{eqnarray}
with $K_\rho$ and $J_\rho$ defined in appendix \ref{sec:kj}
and ${{\Lambda}}^{\nu\rho}_{-}$ in (\ref{eq:lambda}).
Because ${\bar{q}}_{2}^\mu$ is infinitesimally
small we have
\begin{eqnarray}
{}^\star\{{\bar{q}}^\mu(\sigma),{\bar{q}}^\nu(\bar{\sigma})\}&=&
-2\theta^{\mu\alpha}_{0}\partial_\alpha \theta^{\nu\rho}K_\rho(\bar{\sigma},\sigma)
+2\theta^{\nu\alpha}_{0}\partial_\alpha \theta^{\mu\rho}K_\rho(\sigma,\bar{\sigma})
\nonumber\\
&&-\theta^{\mu\alpha}_{0}\partial_{\alpha}{{\Lambda}}^{\nu\rho}_{-}J_\rho(\bar{\sigma},\sigma)
+\theta^{\nu\alpha}_{0}\partial_{\alpha}{{\Lambda}}^{\mu\rho}_{-}J_\rho({\sigma},\bar{\sigma}).
\end{eqnarray}

Using (\ref{eq:ttpm}), (\ref{eq:ij}) and the expression
\begin{equation}
\theta^{\mu\alpha}_{0}\partial_{\alpha}{{\Lambda}}^{\nu\rho}_{-}
=\theta^{\mu\alpha}_{0}\partial_{\alpha}{\theta}^{\nu\rho}
-\theta^{\mu\alpha}_{0}\partial_{\alpha}{{\Lambda}}^{\nu\rho}_{+},
\end{equation}
we obtain
\begin{equation}\label{eq:qbar2}
{}^\star\{{\bar{q}}^\mu(\sigma),{\bar{q}}^\nu(\bar{\sigma})\}=
4I^{\nu\mu}[P(\sigma)]
\theta_{S}(\bar{\sigma},{\sigma})
-4I^{\mu\nu}[P(\bar{\sigma})]
\theta_{S}({\sigma},\bar{\sigma}),
\end{equation}
where $\theta_{S}$ is introduced in (\ref{eq:ths})
and we defined
\begin{equation}\label{eq:iveliko}
I^{\mu\nu}[P]\equiv
\Big{[}
\frac{1}{2}\theta^{\mu\alpha}_{0}\partial_{\alpha}\theta^{\nu\rho}
-\frac{1}{4}\theta^{\nu\alpha}_{0}\partial_{\alpha}{{\Lambda}}_{+}^{\mu\rho}
\Big{]}P_\rho.
\end{equation}

%%%%%%%%%%%%%%%%%%%%%%%%%%%%%%%%%%%%%%%%%%%%%%%%%%%%%%%%%%%%%%%%%%%%%%%%%%%%%%%%
\subsection{The non-commutativity relation}
With the help of (\ref{eq:qbar2})
and the expression
\begin{eqnarray}
{}^\star\{q^\mu(\sigma),{\bar{q}}^\nu(\bar{\sigma})\}&=&
\Big{[}
{{\Lambda}}^{\mu\nu}_{+}[q(\bar{\sigma})]+{{\Lambda}}^{\mu\nu}_{-}[q({\sigma})]
\Big{]}\theta_{S}(\bar{\sigma},\sigma),
\end{eqnarray}
the complete non-commutativity relation
\begin{equation}\label{eq:xx}
{}^\star\{x^\mu(\sigma),{{x}}^\nu(\bar{\sigma})\}=
{}^\star\{q^\mu(\sigma),{\bar{q}}^\nu(\bar{\sigma})\}+
{}^\star\{{\bar{q}}^\mu(\sigma),{q}^\nu(\bar{\sigma})\}+
{}^\star\{{\bar{q}}^\mu(\sigma),{\bar{q}}^\nu(\bar{\sigma})\},
\end{equation}
takes the form
\begin{eqnarray}\label{eq:xxceo}
{}^\star\{x^\mu(\sigma),{{x}}^\nu(\bar{\sigma})\}&=&
\Big{[}
{{\Lambda}}^{\mu\nu}_{+}[q(\bar{\sigma})]
+{{\Lambda}}^{\mu\nu}_{-}[q({\sigma})]
+4I^{\nu\mu}[P(\sigma)]
\Big{]}
\theta_{S}(\bar{\sigma},{\sigma})
\nonumber\\
&-&\Big{[}
{{\Lambda}}^{\nu\mu}_{+}[q({\sigma})]
+{{\Lambda}}^{\nu\mu}_{-}[q({\bar{\sigma}})]
+4I^{\mu\nu}[P(\bar{\sigma})]
\Big{]}
\theta_{S}({\sigma},\bar{\sigma}).
\end{eqnarray}
Using the expression (\ref{eq:ths}) for
the symmetric theta function $\theta_{S}$ we
can rewrite it as
\begin{eqnarray}\label{eq:xxstar}
{}^\star\{x^\mu(\sigma),x^\nu(\bar{\sigma})\}&=&
2\Big{[}
E^{\mu\nu}(\bar{\sigma})-E^{\nu\mu}(\sigma)
\Big{]}\theta(\sigma+\bar{\sigma})
\nonumber\\
&-&2\Big{[}
I^{\mu\nu}(\bar{\sigma})+
I^{\nu\mu}(\sigma)
\Big{]}\theta(\sigma-\bar{\sigma}),
\end{eqnarray}
where we introduced
\begin{equation}\label{eq:omega}
E^{\mu\nu}[q,P]=\frac{1}{2}{{\Lambda}}^{\mu\nu}_{+}[q]-I^{\mu\nu}[P].
\end{equation}

In comparison with the case $b_{\mu\nu}=0$,
the essential difference is the infinitesimally
small and momenta dependent term $I^{\mu\nu}$.
It appears as a multiplier
not only of the function $\theta(\sigma+\bar{\sigma})$,
but of the function $\theta(\sigma-\bar{\sigma})$ as well.
This means that it causes the non-commutativity not only
on the string end-points,
but on the string interior, also.

Using (\ref{eq:lambda}) and the fact that $\theta^{\mu\nu}_{eff}$
is equal to $\Delta\theta$
we can rewrite  the expression (\ref{eq:iveliko}) as
\begin{equation}\label{eq:ivelexp}
I^{\mu\nu}[P]=
\frac{1}{2}
\theta^{\mu\nu\rho}P_\rho
-\frac{1}{2}\theta^{\mu\nu}_{eff}[-\theta_{0}P]
+\frac{1}{4}
\theta^{\nu\alpha}_{0}\partial_\alpha{{\Lambda}}_{-}^{\mu\rho}P_\rho,
\end{equation}
where $\theta^{\mu\nu\rho}$ is defined in
(\ref{eq:tetatrig}).
Similarly, the expression (\ref{eq:omega}) gains the form
\begin{equation}\label{eq:eexp}
E^{\mu\nu}[q,P]=
\frac{1}{2}\theta^{\mu\nu}[q]
-\frac{3}{\kappa}(G^{-1}_{eff})^{\mu\nu}[bq]
+\frac{1}{2}{\theta}_{eff}^{\mu\nu}[-\theta_{0}P]
-\frac{1}{2}\theta^{\mu\nu\rho}P_\rho
-\frac{1}{4}
\theta^{\nu\alpha}_{0}\partial_\alpha{{\Lambda}}_{-}^{\mu\rho}P_\rho.
\end{equation}

%%%%%%%%%%%%%%%%%%%%%%%%%%%%%%%%%%%%%%%%%%%%%%%%
\subsection{Removing the unphysical terms}
If we separate the center of mass variable
\begin{eqnarray}\label{eq:xcm}
x^\mu(\sigma)=X^\mu(\sigma)+x^{\mu}_{cm},
&&
 x^{\mu}_{cm}=\frac{1}{\pi}\int_{0}^{\pi}d\sigma x^\mu(\sigma),
\end{eqnarray}
from the expression (\ref{eq:xxstar}) follows
\begin{eqnarray}\label{eq:xxp}
{}^\star\{X^\mu(\sigma),X^\nu(\bar{\sigma})\}
&=&
2\Big{[}
E^{\mu\nu}(\bar{\sigma})-E^{\nu\mu}({\sigma})
\Big{]}
\Big{[}
\theta(\sigma+\bar{\sigma})-1/2
\Big{]}
\nonumber\\
&+&2I^{\nu\mu}(\sigma)
\Big{[}-\theta(\sigma-\bar{\sigma})
-\frac{1}{2}+\frac{\sigma}{\pi}\Big{]}
\nonumber\\
&+&2I^{\mu\nu}(\bar{\sigma})
\Big{[}-\theta(\sigma-\bar{\sigma})
+\frac{1}{2}-\frac{\bar{\sigma}}{\pi}\Big{]}
\nonumber\\
&+&\frac{2}{\pi}
\Big{[}
{{_{\circ}}I}^{\mu\nu}(\sigma)
-{{_{\circ}}I}^{\nu\mu}(\bar{\sigma})
-{{_{\circ}}I}^{\mu\nu}_{cm}
+{{_{\circ}}I}^{\nu\mu}_{cm}
\Big{]},
\end{eqnarray}
where we introduced notation
\begin{eqnarray}
{{_{\circ}}I}^{\mu\nu}(\sigma)=\int_{0}^{\sigma}d\eta I^{\mu\nu}(\eta),
&&
{{_{\circ}}I}^{\mu\nu}_{cm}=
\frac{1}{\pi}\int_{0}^{\pi}d\sigma {{_{\circ}}I}^{\mu\nu}(\sigma).
\end{eqnarray}
Notice that $\theta(\sigma+\bar{\sigma})-1/2$ is nonzero, only on the
string end-points, and therefore the symmetric part of $E^{\mu\nu}$
disappears.
Separating the symmetric and the antisymmetric part of
$I^{\mu\nu}=[I^{\mu\nu}+I^{\nu\mu}]/2+[I^{\mu\nu}-I^{\nu\mu}]/2\equiv
I^{(\mu\nu)}+I^{[\mu\nu]}$ we obtain
\begin{eqnarray}\label{eq:xxp1}
{}^\star\{X^\mu(\sigma),X^\nu(\bar{\sigma})\}
&=&
\Big{[}
\theta^{\mu\nu}(\bar{\sigma})+\theta^{\mu\nu}({\sigma})
\Big{]}
\Big{[}
\theta(\sigma+\bar{\sigma})-1/2
\Big{]}
\nonumber\\
&+&2I^{[\mu\nu]}(\bar{\sigma})
\Big{[}
1-2\theta_{S}(\sigma,\bar{\sigma})-\frac{\bar{\sigma}}{\pi}
\Big{]}
\nonumber\\
&+&2I^{[\mu\nu]}({\sigma})
\Big{[}
1-2\theta_{S}(\bar{\sigma},\sigma)-\frac{{\sigma}}{\pi}
\Big{]}
\nonumber\\
&+&2I^{(\mu\nu)}(\bar{\sigma})
\Big{[}
1/2+\theta(\bar{\sigma}-\sigma)-\frac{\bar{\sigma}}{\pi}
\Big{]}
\nonumber\\
&-&
2I^{(\mu\nu)}({\sigma})
\Big{[}
1/2+\theta(\sigma-\bar{\sigma})-\frac{{\sigma}}{\pi}
\Big{]}
\nonumber\\
&+&\frac{2}{\pi}
\Big{[}
{{_{\circ}}I}^{\mu\nu}(\sigma)
-{{_{\circ}}I}^{\nu\mu}(\bar{\sigma})
-{{_{\circ}}I}^{\mu\nu}_{cm}
+{{_{\circ}}I}^{\nu\mu}_{cm}
\Big{]},
\end{eqnarray}
where $\theta_{S}$ is defined in (\ref{eq:ths}).
Notice that $I^{\mu\nu}[P]$ is linear in $P$,
so we can rewrite the above expression as
\begin{eqnarray}\label{eq:xxcom}
{}^\star\{X^\mu(\sigma),X^\nu(\bar{\sigma})\}
&=&
\Big{[}
\theta^{\mu\nu}[q(\bar{\sigma})]
+\theta^{\mu\nu}[q({\sigma})]
\Big{]}
\Big{[}
\theta(\sigma+\bar{\sigma})-1/2
\Big{]}
\nonumber\\
&+&2I^{[\mu\nu]}[P^{a}(\sigma,\bar{\sigma})]
+2I^{(\mu\nu)}[P^{s}(\sigma,\bar{\sigma})],
\end{eqnarray}
with
\begin{eqnarray}\label{eq:ivelsa}
I^{[\mu\nu]}[P]&=&
\frac{3}{8}
\theta^{\mu\nu\rho}P_\rho
-\frac{3}{8}\theta^{\mu\nu}_{eff}[-\theta_{0}P]
+\frac{3}{4\kappa}
\Big{[}
\theta^{\nu\rho}_{0}
\partial_\rho{({G^{-1}_{E}})}^{\mu\sigma}
-
\theta^{\mu\rho}_{0}
\partial_\rho{({G^{-1}_{E}})}^{\nu\sigma}
\Big{]}
P_\sigma,
\nonumber\\
I^{(\mu\nu)}[P]&=&
\frac{1}{8}
\Big{[}
\theta^{\nu\rho}_{0}
\partial_\rho{{\Lambda}}_{-}^{\mu\sigma}
+
\theta^{\mu\rho}_{0}
\partial_\rho{{\Lambda}}_{-}^{\nu\sigma}
\Big{]}P_\sigma,
\end{eqnarray}
and
\begin{eqnarray}\label{eq:pas}
P^{a}_{\mu}(\sigma,\bar{\sigma})&=&
P_{\mu}(\bar{\sigma})\Big{[}
1-2\theta_{S}(\sigma,\bar{\sigma})-\frac{\bar{\sigma}}{\pi}
\Big{]}
+P_{\mu}({\sigma})
\Big{[}
1-2\theta_{S}(\bar{\sigma},\sigma)-\frac{{\sigma}}{\pi}
\Big{]}
\nonumber\\
&+&\frac{1}{\pi}\Big{[}{_{0}}P_\mu(\sigma)
+{{_{0}}P_\mu(\bar{\sigma})}\Big{]}
-\frac{2}{\pi}\ {{_{0}}P_{\mu}^{cm}},
\nonumber\\
P^{s}_{\mu}(\sigma,\bar{\sigma})&=&
P_{\mu}(\bar{\sigma})
\Big{[}
\frac{1}{2}
+\theta(\bar{\sigma}-\sigma)-\frac{\bar{\sigma}}{\pi}
\Big{]}
-P_{\mu}({\sigma})
\Big{[}
\frac{1}{2}
+\theta(\sigma-\bar{\sigma})-\frac{{\sigma}}{\pi}
\Big{]}
\nonumber\\
&+&\frac{1}{\pi}\Big{[}{_{0}}P_\mu(\sigma)
-{{_{0}}P_\mu(\bar{\sigma})}\Big{]}.
\end{eqnarray}
We used the useful notation
\begin{eqnarray}
{_\circ}{P}_{\mu}(\sigma)=\int_{0}^{\sigma}d\eta P_\mu(\eta),
&&
{_\circ}P_{\mu}^{cm}=\frac{1}{\pi}\int_{0}^{\pi}d\eta\  {_\circ}P_\mu(\eta),
\nonumber\\
{^\circ}{P}_{\mu}(\sigma)=\int_{\sigma}^{\pi}d\eta P_\mu(\eta),
&&
{^\circ}P_{\mu}^{cm}=\frac{1}{\pi}\int_{0}^{\pi}d\eta\  {^\circ}P_\mu(\eta).
\end{eqnarray}

%%%%%%%%%%%%%%%%%%%%%%%%%%%%%%%%%%%%%%%%%%%%%%%%%%%%%%%%%%%%%%%%%%%%%%
\subsection{Momenta dependent non-commutativity parameters}
Let us first consider the case $\sigma=\bar{\sigma}$.
The eq. (\ref{eq:pas}) reduces to
\begin{eqnarray}
P^{a}_{\mu}(\sigma,\sigma)&=&
2P_{\mu}(\sigma)\Big{[}1-2\theta_{S}(\sigma,\sigma)-
\frac{\sigma}{\pi}\Big{]}
+\frac{2}{\pi}\Big{[}{_{0}P_\mu}(\sigma)-{_{0}P_{\mu}^{cm}}\Big{]},
\nonumber\\
P^{s}_{\mu}(\sigma,\sigma)&=&0,
\end{eqnarray}
and the eq.(\ref{eq:xxcom}) becomes
\begin{eqnarray}
{}^\star\{X^\mu(\sigma),X^\nu({\sigma})\}
&=&
2\theta^{\mu\nu}({\sigma})
\Big{[}\theta(2\sigma)-1/2\Big{]}
+2I^{[\mu\nu]}[P^{a}(\sigma,\sigma)].
\end{eqnarray}
Taking into account that $P_\mu(0)={_{0}P_\mu}(0)=P_\mu(\pi)=0$
and
${_\circ}P_{\mu}^{cm}+{^\circ}P_{\mu}^{cm}=\pi P_{\mu}^{cm}=
{_{0}P}_\mu(\pi)$
we obtain
\begin{eqnarray}
P^{a}_{\mu}(0,0)=
-\frac{2}{\pi}\ {_{0}P^{cm}_{\mu}},
&&
P^{a}_{\mu}(\pi,\pi)=
\frac{2}{\pi}\ {^{0}P^{cm}_{\mu}}.
\end{eqnarray}
The commutation relations on the string endpoints are therefore equal to
\begin{equation}\label{eq:xxoexp}
{}^\star\{X^\mu(0),X^\nu({0})\}
= -\theta^{\mu\nu}[q(0)]
-\frac{4}{\pi}I^{[\mu\nu]}[{_{0}P_{cm}}],
\end{equation}
\begin{equation}\label{eq:xxpexp}
{}^\star\{X^\mu(\pi),X^\nu({\pi})\}
=\theta^{\mu\nu}[q(\pi)]
+\frac{4}{\pi}I^{[\mu\nu]}[{^{0}P_{cm}}],
\end{equation}
or explicitly
\begin{eqnarray}\label{eq:xxnula}
{}^\star\{X^\mu(0),X^\nu({0})\}
&=& -\theta^{\mu\nu}[q(0)]
-\frac{3}{2\pi}
\theta^{\mu\nu\rho}{_{0}P^{cm}_\rho}
+\frac{3}{2\pi}\theta^{\mu\nu}_{eff}[-\theta_{0}\ {_{0}P^{cm}}]
\nonumber\\
&-&\frac{3}{\pi\kappa}
\Big{[}
\theta^{\nu\rho}_{0}
\partial_\rho{({G^{-1}_{E}})}^{\mu\sigma}
-
\theta^{\mu\rho}_{0}
\partial_\rho{({G^{-1}_{E}})}^{\nu\sigma}
\Big{]}
{_{0}P^{cm}_\sigma},
\end{eqnarray}
\begin{eqnarray}\label{eq:xxpi}
{}^\star\{X^\mu(\pi),X^\nu({\pi})\}
&=&\theta^{\mu\nu}[q(\pi)]
+\frac{3}{2\pi}
\theta^{\mu\nu\rho}{^{0}P^{cm}_\rho}
-\frac{3}{2\pi}\theta^{\mu\nu}_{eff}[-\theta_{0}{^{0}P^{cm}}]
\nonumber\\
&+&\frac{3}{\pi\kappa}
\Big{[}
\theta^{\nu\rho}_{0}
\partial_\rho{({G^{-1}_{E}})}^{\mu\sigma}
-
\theta^{\mu\rho}_{0}
\partial_\rho{({G^{-1}_{E}})}^{\nu\sigma}
\Big{]}
{^{0}P^{cm}_\sigma}.
\end{eqnarray}
The term $\theta^{\mu\nu}[q]$ is standard one
and has been obtained in Refs. \cite{CSS,HKK,ARS,HY,DS}.
It depends on the effective coordinate $q^\mu$
and it is nontrivial only on the string end-points.
Because of the boundary condition
$\bar{q}^\mu(0)=\bar{q}^\mu(\pi)=0$,
we can formally rewrite this term as $x^\mu$
dependent i.e. like $\theta^{\mu\nu}[x(0)]$ and
$\theta^{\mu\nu}[x(\pi)]$.

The other terms
are infinitesimally small and momenta dependent.
Their contribution is nontrivial,
not only on the string end-points,
but on the string interior, also.

The part of the second term
has been obtained in Ref. \cite{HY},
but only on the boundary.
In fact, using the approximation from that article,
that the coordinate is linear in $\sigma$
we have
\begin{eqnarray}
{_\circ}P_{\mu}^{cm}=\frac{\pi^{2}}{6}
p_{\mu}(0),
&&
{^\circ}P_{\mu}^{cm}=\frac{\pi^{2}}{6}
p_{\mu}(\pi),
\end{eqnarray}
so that second term at string endpoints produces
\begin{eqnarray}\label{eq:hy}
-\frac{3}{2\pi}\theta^{\mu\nu\rho}{{_\circ}{P}}_{\rho}^{cm}
=-\frac{\pi}{4}\theta^{\mu\nu\rho}p_\rho(0),
&&
\frac{3}{2\pi}\theta^{\mu\nu\rho}{{^\circ}P_{\rho}^{cm}}
=\frac{\pi}{4}\theta^{\mu\nu\rho}p_\rho(\pi),
\end{eqnarray}
which are the terms obtained in \cite{HY}.
In our case
the non-commutativity parameter contains
additional infinite number of terms
with
$\partial^{n+1}_{\sigma}P_{\mu}(\sigma)
\Big{|}_{\sigma=0,\pi},\quad (n \geq 1)$
on both string endpoints.
Note that the coefficients in (\ref{eq:hy}) comes from
 the two terms of (\ref{eq:ivelexp}). Besides the explicit $1/2$,
in addition we have
$-1/8$ from the anti-symmetrized third term, which
produces our $3/8$ in (\ref{eq:ivelsa}).
The coefficient in the \cite{HY} is just $1/2$.
So, the third term of (\ref{eq:ivelexp}) has been missed under
approximations of the Ref. \cite{HY}.

The other terms in (\ref{eq:xxnula}) and (\ref{eq:xxpi})
are our improvement
and they have not been obtained in the literature before.
They are infinitesimally small and
appear both at the string end-points and
at the string interior.
The third one $\theta^{\mu\nu}_{eff}[\theta_{0}\,{_{0}P^{cm}}]$
depends on the effective background fields
$G^{\mu\nu}_{eff}[q]$ and
$B_{\mu\nu}^{eff}[\theta_{0}\,{_{0}P^{cm}}]$
exactly in the same way as the standard
non-commutativity parameter  $\theta^{\mu\nu}$
depends on the initial background fields
$G_{\mu\nu}$ and $B_{\mu\nu}[q]$.
In general,
it could depend both on
$q^\mu$ (through $G^{eff}_{\mu\nu}$)
and on $_{0}P^{cm}_\mu$ (through $B^{eff}_{\mu\nu}$).
Since $B^{eff}_{\mu\nu}$ is infinitesimal,
in our particular case of weakly curved background,
only constant term of $G^{eff}_{\mu\nu}$ survives.
In fact
$\theta^{\mu\nu}_{eff}=$
$\Delta\theta^{\mu\nu}$
and we can include it in the first term and obtain
\begin{eqnarray}
\theta^{\mu\nu}[q+\frac{3}{2\pi}\theta_{0}\,
{_{0}P^{cm}}]\Big{|}_{\sigma=0},
&&
\theta^{\mu\nu}[q+\frac{3}{2\pi}\theta_{0}\,
{^{0}P^{cm}}]\Big{|}_{\sigma=\pi}.
\end{eqnarray}

The last term contains the derivatives of the effective
metric,
which are constant as we are working with the weakly
curved background.

The non-commutativity relation on the string interior
is equal to
\begin{equation}
{}^\star\{X^\mu(\sigma),X^\nu(\bar{\sigma})\}
=
2I^{[\mu\nu]}[P^{a}(\sigma,\bar{\sigma})]
+2I^{(\mu\nu)}[P^{s}(\sigma,\bar{\sigma})],
\end{equation}
with
\begin{eqnarray}
P^{a}_\mu(\sigma,\sigma)&=&2P_\mu(\sigma)\Big{[}
\frac{1}{2}
-\frac{\sigma}{\pi}\Big{]}
+\frac{2}{\pi}\Big{[}{_{0}P_\mu}(\sigma)-{_{0}P_\mu^{cm}}\Big{]},
\nonumber\\
P^{s}_\mu(\sigma,\sigma)&=&0,
\end{eqnarray}
for $\sigma=\bar{\sigma}\neq{0,\pi}$,
\begin{eqnarray}
P^{a}_{\mu}(\sigma,\bar{\sigma})&=&
P_{\mu}({\sigma})
-\frac{1}{\pi}
\Big{[}
{\bar{\sigma}}P_{\mu}(\bar{\sigma})
+{{\sigma}}P_{\mu}({\sigma})
\Big{]}
\nonumber\\
&+&\frac{1}{\pi}\Big{[}{_{0}}P_\mu(\sigma)
+{{_{0}}P_\mu(\bar{\sigma})}\Big{]}
-\frac{2}{\pi}\ {{_{0}}P_{\mu}^{cm}},
\nonumber\\
P^{s}_{\mu}(\sigma,\bar{\sigma})&=&
-P_{\mu}({\sigma})
-\frac{1}{\pi}
\Big{[}
\bar{\sigma}P_\mu(\bar{\sigma})-
{\sigma}P_\mu({\sigma})
\Big{]}
\nonumber\\
&+&\frac{1}{\pi}\Big{[}{_{0}}P_\mu(\sigma)
-{{_{0}}P_\mu(\bar{\sigma})}\Big{]},
\end{eqnarray}
for $\sigma>\bar{\sigma}$ and
\begin{eqnarray}
P^{a}_{\mu}(\sigma,\bar{\sigma})&=&
P_{\mu}(\bar{\sigma})
-\frac{1}{\pi}
\Big{[}
{\bar{\sigma}}P_{\mu}(\bar{\sigma})
+{{\sigma}}P_{\mu}({\sigma})
\Big{]}
\nonumber\\
&+&\frac{1}{\pi}\Big{[}{_{0}}P_\mu(\sigma)
+{{_{0}}P_\mu(\bar{\sigma})}\Big{]}
-\frac{2}{\pi}\ {{_{0}}P_{\mu}^{cm}},
\nonumber\\
P^{s}_{\mu}(\sigma,\bar{\sigma})&=&
P_{\mu}(\bar{\sigma})
-\frac{1}{\pi}
\Big{[}
\bar{\sigma}P_\mu(\bar{\sigma})-
{\sigma}P_\mu({\sigma})
\Big{]}
\nonumber\\
&+&\frac{1}{\pi}\Big{[}{_{0}}P_\mu(\sigma)
-{{_{0}}P_\mu(\bar{\sigma})}\Big{]},
\end{eqnarray}
for $\bar{\sigma}>\sigma$.
Note that the contribution to the string
interior comes from
the infinitesimal term
$I^{\mu\nu}$ defined in (\ref{eq:iveliko}).

Keeping in mind (\ref{eq:pdual}),
we conclude
that the non-commutative parameters depend
both on the effective coordinate $q$
and its T-dual $\tilde{q}$.
%%%%%%%%%%%%%%%%%%%%%%%%%%%%%%%%%%%%%%%%%%%%%%%%%%%%%%%%%%%%%%%%%%%%%%%%%%%%%%%%%%%%%%%%%%%%%%%%%%
\subsection{Canonical quantization}
Let us shortly comment the canonical quantization of the considered theory.
We will discuss two possible rules of associating
the operators to the variables and
replacement of the brackets with the commutator.

First of all,
the straightforward generalization of the standard quantization procedure is
not possible.
There appears the problem, because the commutator
of the coordinates of the string end-points
does not close on coordinates but
depends on the momenta, also.
So, generally we should consider the operator functions
of coordinates and momenta $\hat{f}(\hat{x},\hat{\pi})$
and normal ordering among both $x$'s and $x$ and $\pi$.

There is the other possibility, we suggested in \cite{DS}.
One can consider the effective variables
$q^\mu$ and $p_\mu$
as fundamental variables and introduce some
normal ordering $::$ for corresponding operators
${\hat{q}}^\mu$ and ${\hat{p}}_\mu$.
We can rewrite (\ref{eq:xpres}) in the operator form,
\begin{eqnarray}
{\hat{x}}^\mu(\sigma)&=&
{\hat{q}}^\mu(\sigma)
-\int_{0}^{\sigma}d\eta
:\Big{[}\theta^{\mu\nu}[{\hat{q}}(\eta)]{\hat{p}}_\nu(\eta)
+\frac{1}{2}{\Lambda}^{\prime\mu\nu}_{-}
[{\hat{q}}(\eta)]{\hat{P}}_\nu(\eta)\Big{]}:,
\nonumber\\
{\hat{\pi}}_\mu(\sigma)&=&
2{\hat{p}}_\mu(\sigma)
+\Big{[}Gb^{-1}\beta[-\theta_{0}\hat{P}]
g^{-1}\Big{]}_{\mu}^{\ \nu}{\hat{p}}_\nu.
\end{eqnarray}
Note that
because $\hat{x}^\mu$ depends both on $\hat{q}^\mu$ and $\hat{p}_\mu$,
we used the normal ordering to define it,
but we did not need it for ${\hat{\pi}}_\mu$ because it
is only $\hat{p}_\mu$ dependent.
Now, it is possible
to assign the operator to any function $f(x,\pi)$
\begin{eqnarray}
f(x,\pi)&\rightarrow&
:\hat{f}
\Big{\{}
{\hat{q}}^\mu-\int_{0}^{\sigma}d\sigma_{0}
\Big{[}
\theta^{\mu\nu}[\hat{q}]{\hat{p}}_\nu
+\frac{1}{2}{\Lambda}^{\prime\mu\nu}_{-}[\hat{q}]{\hat{P}}_\nu
\Big{]},
\nonumber\\
&&
 2{\hat{p}}_\mu
+\Big{[}Gb^{-1}\beta[-\theta_{0}\hat{P}]
g^{-1}\Big{]}_{\mu}^{\ \nu}{\hat{p}}_\nu
\Big{\}}:,
\end{eqnarray}
which we will consider as an operator of the basic
variables $:\hat{f}(\hat{q},\hat{p}):$.
The new star product $\star$,
associated with the normal ordering $::$,
can be defined
demanding the prescription
$f\star g\rightarrow \hat{f}\hat{g}$.
It is defined along the whole string
and its properties and its relation
with the Moyal and the Kontsevich product \cite{K}
could be matter of further investigations.
It would also be a challenge to construct the field theories
on such non-commutative spaces.
%%%%%%%%%%%%%%%%%%%%%%%%%%%%%%%%%%%%%%%%%%%%%%%%%%%%%%%%%%%%%%%%%%%%%%%%%%%%%%%%%%%%%%%%%%%%%%%%%%%%%%%%%%%%%%%%%%5
\section{Conclusion}
\cleq
We investigated the propagation of the open string
in the background with  the constant
metric tensor $G_{\mu\nu}$ and
the Kalb-Ramond field linear in coordinate
$B_{\mu\nu}[x]=b_{\mu\nu}+\frac{1}{3}B_{\mu\nu\rho}x^\rho$.
We considered the case of the infinitesimally small
$B_{\mu\nu\rho}$ and did all the calculations up to its first order.
This, so-called {\it weakly curved} background is the solution of the space-time equations of motion,
where the Ricci tensor $R_{\mu\nu}$ was neglected as the infinitesimally small variable of the second order.

We treated the
boundary conditions at the open string end-points as constraints.
We used the modified canonical method
developed in \cite{DS},
where the constant background term $b_{\mu\nu}$ was zero.
After nontrivial calculations we obtained the complete
set of the constraints in a compact form.
We showed that they are of the second class and we solved them explicitly.

On the solution of the constraints we obtained the
effective theory. It is
an unoriented closed string theory on the orbifold,
defined
as an $\Omega$-even ($\sigma$-parity transformation) projection of
the initial one.
The effective string propagates in the effective background (see (\ref{eq:gbefff}) and (\ref{eq:xgb})).
The effective metric formally has the same form as in the case of the constant background,
but here it is coordinate dependent due to the coordinate dependence of the initial
Kalb-Ramond field.

The term with the effective Kalb-Ramond field is of greater importance.
It is generally accepted that
the fact that the theory is $\Omega$-even forces the effective
Kalb-Ramond field to vanish.
The standard proof of this statement,
contains the assumption that the Kalb-Ramond field depends on the
effective coordinate $q^\mu$.
But unexpectedly, we obtain the infinitely small
effective Kalb-Ramond field,
which depends on the effective momentum $B^{eff}_{\mu\nu}[-\theta_{0}P]$
in the Hamiltonian approach and
on the $\sigma$-integral of the $\tau$-derivative
of the effective coordinate
$B^{eff}_{\mu\nu}[2b\dot{Q}]$
in the Lagrangian approach.
In both cases its argument is $\Omega$-odd variable.
This introduces the additional minus sign
under $\Omega$-transformation
of $B_{\mu\nu}^{eff}$,
which brakes the proof
that it should vanish.
As was shown in subsection \ref{sec:effkr},
in both Lagrangian and Hamiltonian approaches,
the effective Kalb-Ramond field has a form
$B_{\mu\nu}^{eff}[2b\tilde{q}]$
where $\tilde{q}^\mu$ is
T-dual effective coordinate.

As well as in the previous investigations by
canonical methods, the expression of the initial coordinate $x^\mu$
on the solution of the boundary condition (\ref{eq:xpres}) depends
not only on the effective coordinates
but also on
the effective momenta which
is a source of the non-commutativity.
In our particular case this dependence is nonlinear,
which made the calculation
much more complicated.

The essential difference between the non-commutativity parameters
for $b_{\mu\nu}\neq0$ and $b_{\mu\nu}=0$,
is caused by the fact that
the term $\theta^{\mu\nu}[q(\eta)]$
in the expression (\ref{eq:xpres}) for $x^\mu$ is
no longer infinitesimal,
but contains the finite term $\theta^{\mu\nu}_{0}$.
Therefore,
the star bracket ${}^\star\{{\bar{q}}^\mu,{\bar{q}}^\nu\}$
which was an infinitesimal of the second order in \cite{DS},
becomes an infinitesimal of the first order
and can not be neglected.

The complete non-commutativity relation is given by (\ref{eq:xxstar}).
Beside two coordinate dependent terms,
we obtain terms which depend on
$P_\mu=\kappa g_{\mu\nu}{\tilde{q}}^\nu$,
as can be seen from the expressions (\ref{eq:eexp}).
Separating the center of mass of the coordinate $x^\mu$,
the non-commutativity relation transfers to (\ref{eq:xxp1}).
It turns out that the non-commutativity parameter
on the string endpoints depends
only on terms antisymmetric in $\mu,\nu$.
So, the one of the two coordinate dependent terms
$G^{-1}_{eff}[bq]$ does not contribute.
The other $\theta^{\mu\nu}[q]$ is the standard one and
exists only at the string end-points.
All the other contributions are from the
infinitesimally small and
momenta dependent term $I^{\mu\nu}$ given by (\ref{eq:ivelexp}).
They are nontrivial at both the string boundary and
its interior.

Under some approximations including low energy limit and assumption that $x^\mu$
depends linearly on $\sigma$,
the first term in $I^{\mu\nu}$,
$\theta^{\mu\nu}_{\ \ \rho}{\bar{q}_{0}}^\rho=\theta^{\mu\nu\rho}P_\rho$
on the string end-points produces
that obtained in the Ref.\cite{HY}.
We find that its contribution is nontrivial in the string interior as well.

The second term
\begin{eqnarray}
{{\theta}}_{eff}^{\mu\nu}&=&
-\frac{2}{\kappa}
\Big{[}
G^{-1}_{E}(G^{eff},B^{eff})B^{eff}G^{-1}_{eff}
\Big{]}^{\mu\nu},
\end{eqnarray}
has never been obtained in the literature before.
By definition it depends on the effective background fields
$G^{eff}_{\mu\nu}$ and $B^{eff}_{\mu\nu}$
in the same way as the standard non-commutativity parameter $\theta^{\mu\nu}$ depends on
the initial background fields
$G_{\mu\nu}$ and $B_{\mu\nu}$.
Note that, the initial
metric tensor is constant and
the initial Kalb-Ramond field depends on coordinate,
so that the standard non-commutative parameter
$\theta^{\mu\nu}$ depends on coordinate.
On the other hand,
the effective metric depends on the effective coordinate,
${G}^{eff}_{\mu\nu}[q]$, and the effective Kalb-Ramond field depends on
the effective
momentum ${B}^{eff}_{\mu\nu}[-\theta_{0}P]$.
So, the new term of the non-commutativity parameter
could in general
depend both on the coordinate and the momentum.
In the case of the weakly curved background,
${B}^{eff}_{\mu\nu}$ is infinitesimally small and
the above expression turns to
\begin{eqnarray}
{{\theta}}_{eff}^{\mu\nu}[-\theta_{0}P]&=&
-\frac{2}{\kappa}
\Big{[}g^{-1}{{B}^{eff}}[-\theta_{0}P]g^{-1}\Big{]}^{\mu\nu},
\end{eqnarray}
and consequently it depends only on the effective momentum.
One can expect, that in higher order in $B_{\mu\nu\rho}$
this new term will depend both on effective coordinate and momentum.

The third term depends on the effective momentum
with the coefficients proportional
to the derivative of the effective metric.

Because the integral of the effective momentum is proportional to
the T-dual effective coordinate $P_\mu=\kappa g_{\mu\nu}{\tilde{q}}^\nu$,
we can conclude that all background fields depend on the effective
coordinate $q^\mu$ and its T-dual ${\tilde{q}}^\mu$.
In comparison with the case $b_{\mu\nu}=0$,
all the improvements are
${\tilde{q}}^\mu$-dependent and infinitesimally small.

We obtained the complete set of the noncommutativity parameters,
for the case of the of the weakly curved background when
the field strength of the Kalb-Ramond field is infinitesimal parameter.
We proved that the momentum dependent term of the Ref.\cite{HY}
really exists and we found the new momentum dependent terms.
We offered the new interpretation, in which the non-commutativity
parameters depend on the effective coordinate $q^\mu$
and its T-dual ${\tilde{q}}^\mu$.

%%%%%%%%%%%%%%%%%%%%%%%%%%%%%%%%%%%%%%%%%%%%%%%%%%%%%%%%%%%%%%%%%%%%%%%%%%%%%%%%%%%%%%%%%%%%%%%%%%%
Let us bring some arguments in support of the regularization independence of our result.
It is known in the literature,
that in the vicinity of the open string boundary,
some regularization should be imposed in order
to escape singularity, even for $B_{\mu\nu}=const$
(for more details see Ref. \cite{SJS}).

In our approach,
the transition from the infinite set of constraints
$\gamma^{n}_{\mu}\Big{|}_{0}$ at the point
to the $\sigma$-dependent constraints
$\Gamma^{S}_{\mu}(\sigma)$ and $\Gamma^{A}_{\mu}(\sigma)$
defined at (\ref{eq:gamasa}),
is some sort of regularization.
In fact Poisson bracket between $\sigma$-dependent constraints
(\ref{eq:poisgama}) is well defined,
while the brackets between the coefficient of its Taylor expansion
$\gamma^{n}_{\mu}\Big{|}_{0}=
\partial^{n}_{\sigma}
\Gamma_{\mu}(\sigma)\Big{|}_{\sigma=0}$
is formally proportional to the derivative of the $\delta$-function
at zero,
which is not well defined.

The arbitrariness of our regularization is built into the
$\sigma$-dependence. So, we can change the regularization choosing
some new parametrization $\tilde{\sigma}=\tilde{\sigma}(\sigma)$
which preserves end-points, $\tilde{\sigma}(0)=0$ and
$\tilde{\sigma}(\pi)=\pi$. Therefore, we can expect that the
non-commutativity relations at the boundary (\ref{eq:xxnula}) and
(\ref{eq:xxpi}) are not regularization dependent and that they
describe the physical phenomenon. We expect that the other kinds
of regularization "discretizating`` the string \cite{SJS} or
using the limit of 2d world-sheet theory two-point functions
\cite{SW} will lead to the same result for the non-commutativity
of the string end-points.

We will also comment the influence of the non-locality of
the present theory on the low energy field theory of the open strings
living on Dp-brane which we will call generalized
Dirac-Born-Infeld (DBI) theory. Note that in the flat background,
for $B_{\mu\nu}=const$, the corresponding term in the action is
quadratic in the dynamical variables. So, it is possible to treat
it as an interaction, or as a kinetic part. The first approach
produces the DBI Lagrangian as the effective Lagrangian for the
slowly varying fields \cite{FT}. In the second approach, the canonical
analysis, similar to that of the present paper, but for
$B_{\mu\nu\rho}=0$, could be performed. Then, the Dp-brane
manifold turns to the non-commutative one, with constant
non-commutative parameter $\theta_{0}^{\mu\nu}$ (\ref{eq:gtetao}) and the
effective action contains only the term with the constant part of
the effective metric $g_{\mu\nu}$ (\ref{eq:gtetao}). The non-commutative (NC)
DBI action expressed in terms of the open string variables
$g_{\mu\nu}$ and $\theta_{0}^{\mu\nu}$ is obtained. Note that
because the effective action depends only on  $g_{\mu\nu}$, the
$\theta_{0}^{\mu\nu}$-dependence appears only in the star product
\cite{SW}. Let us stress that in all calculations the central
assumption is that the derivatives of the antisymmetric fields do
not appear, so that the path integral is Gaussian \cite{FT}.

In our case of the weakly curved background,
the Kalb-Ramond field, besides the constant part $b_{\mu\nu}$
contains also the linear one $\frac{1}{3}B_{\mu\nu\rho}x^\rho$.
Therefore, the corresponding terms in the action are quadratic
($S_{b}=\epsilon^{\alpha\beta}b_{\mu\nu}\int d^2\xi \partial_\alpha x^\mu
\partial_\beta x^\nu$)
and of the third power in $x^\mu$
($S_{3}=\frac{\epsilon^{\alpha\beta}}{3}B_{\mu\nu\rho}
\int d^2\xi \partial_\alpha x^\mu
\partial_\beta x^\nu x^\rho$).
As before, the one possibility is to treat both terms $S_{b}$ and $S_{3}$
as the interaction.
This will produce the commutative theory,
with the low energy field theory
being some extension of the DBI action,
originated in $S_{3}$.
There is also the technical problem, due to the fact that $S_{3}$
is not quadratic in coordinates and consequently the path integral is not Gaussian.

The second possibility is to treat the quadratic part $S_{b}$
as the kinetic term and $S_{3}$ as interaction.
Then, the canonical analysis which includes $S_b$ but not $S_3$ is relevant.
It gives the constant non-commutativity parameter $\theta^{\mu\nu}_{0}$,
which produces non-commutativity with Moyal star product.
There is only one term in the effective action with $g_{\mu\nu}$
so that $\theta^{\mu\nu}_{0}$ appears only in the star product.
But, the perturbation $S_{3}$ will change NC DBI action and make the path integral non Gaussian.

The third possibility is to treat both $S_{b}$ and $S_{3}$
as "kinetic terms".
At first sight, this seams unusual because $S_{3}$
contains third power in $x^\mu$.
But,
considering $S_{3}$ as perturbation,
in the present paper we found the new commutation relations (\ref{eq:xxcom})
and new effective action (\ref{eq:acteff}).

In our formulation, the expectation value of the Wilson loop
would formally have  the form
\begin{equation}\label{eq:EVWloop}
\langle W(A) \rangle  \sim \int Dp Dq e^{-i \int_\Sigma d^2\xi[p {\dot q}
- {\cal{H}}_{c}^{eff}(q,p)] } W_\star (A)  .
\end{equation}
Because the vector field $A_\mu$
leaves on the boundary, the Wilson loop
\begin{equation}\label{eq:Wloop}
W_\star (A) = e_\star^{-i \int_{\partial \Sigma} d x^\mu A_\mu (x)} ,
\end{equation}
is defined with respect to the star product $\star$,
associated with the non-commutativity relations
on the boundary
(\ref{eq:xxnula})
and (\ref{eq:xxpi}).

There are three important differences in comparison with the previous cases.
First,
the effective  Hamiltonian obtains non-trivial term,
with antisymmetric open string variable
$B^{eff}_{\mu\nu}=-\frac{\kappa}{2}(g\Delta\theta g)_{\mu\nu}$,
which is essentially infinitesimal,
momentum dependent improvement of the non-commutativity parameter (\ref{eq:deltag}).
Second,
the non-commutativity is not limited to the string end-points,
but it is extended to the whole string.
Third,
even at the string boundary
the non-commutativity parameter is not constant,
but beside the expected coordinate dependant parts, acquires
the unexpected momentum dependent parts.

From the first difference, it is clear that in the case of the weakly curved background,
the $\theta$-dependence does not appear only in the star product.
It also appears in the low energy field theory
(the extension of the NC DBI theory)
through $B^{eff}_{\mu\nu}$.

The second difference could potentially make remarkable changes.
Following Ref. \cite{FT}, the path integral in (\ref{eq:EVWloop}) is
evaluated by integrating first over all internal points of
$\Sigma$, which reduces the integration to the one over the
boundary $\partial \Sigma$. Only the first term $e^{-i \int_\Sigma
d^2\xi[p {\dot q} - {\cal{H}}_{c}^{eff}(q,p)] }$ depends on the
internal points and it is defined with the ordinary product and
not with the star one. Therefore, we expect that in our case of
the weakly curved background, non-commutativity of the string
interior will not influence on the form of the DBI action.

The third difference, will essentially change the star product,
turning the constant $\theta_0^{\mu\nu}$ to the both coordinate
and momentum dependent non-commutativity parameter
(\ref{eq:xxnula})
and (\ref{eq:xxpi}). This product is not the associative Moyal one,
but is some
generalization of the non-associative Kontsevich one (see section
6. and Ref.\cite{CSS}).

Consequently, we expect that in the case of the weakly curved
background, the space-time low energy effective action should be
expressed on the non-commutative space-time defined by (\ref{eq:xxnula})
and (\ref{eq:xxpi}) as generalized NC DBI action acquiring two infinitesimal
improvements: one explicit trough the effective Kalb-Ramond field
$B^{eff}_{\mu\nu}$ defined in (\ref{eq:gbtoeff}) and the other through the new
non-commutativity parameter with the additional non-constant
terms.

There exist the serious technical problems in obtaining the
generalized NC DBI action, because the path integral is not
Gaussian. The fact that non-Gaussian part is infinitesimal gives
hope that the problem can be solved perturbativelly.
%%%%%%%%%%%%%%%%%%%%%%%%%%%%%%%%%%%%%%%%%%%%%%%%%%%%%%%%%%%%%%%%%%%%%%%%%%%%%%%%%%%%%%%%%%%%
\appendix
\section{Step and delta functions}\label{sec:sdf}
\cleq
The delta and the step functions, symmetric and antisymmetric under $\sigma$-parity,
at the interval $\sigma\in[-\pi,\pi]$,
have the form
\begin{eqnarray}\label{eq:dsa}
\delta_{S}(\sigma,\bar{\sigma})&=&
\frac{1}{2}[\delta(\sigma-\bar{\sigma})+\delta(\sigma+\bar{\sigma})]
=\frac{1}{2\pi}\Big{[}1+2\sum_{n\geq{1}}\cos{n\sigma}\cos{n\bar{\sigma}}\Big{]},
\nonumber\\
\delta_{A}(\sigma,\bar{\sigma})&=&
\frac{1}{2}[\delta(\sigma-\bar{\sigma})-\delta(\sigma+\bar{\sigma})]
=-\frac{1}{2\pi}\Big{[}{\sigma}
+2\sum_{n\geq{1}}\frac{1}{n}\cos{n\sigma}\sin{n\bar{\sigma}}\Big{]},
\end{eqnarray}
\begin{eqnarray}\label{eq:ths}
\theta_{S}(\sigma,\bar{\sigma})&\equiv&
\frac{1}{2}[\theta(\sigma-\bar{\sigma})+\theta(\sigma+\bar{\sigma})]
=\frac{1}{2\pi}\Big{[}\sigma
+2\sum_{n\geq{1}}\frac{1}{n}\sin{n\sigma}\cos{n\bar{\sigma}}\Big{]},
\nonumber\\
\theta_{A}(\sigma,\bar{\sigma})&\equiv&
\frac{1}{2}[\theta(\sigma-\bar{\sigma})-\theta(\sigma+\bar{\sigma})]=
-\frac{1}{2\pi}\Big{[}\bar{\sigma}
+2\sum_{n\geq{1}}\frac{1}{n}\cos{n\sigma}\sin{n\bar{\sigma}}\Big{]}.
\end{eqnarray}

We used the following properties for symmetric $f_{S}(-{\sigma})=f_{S}({\sigma})$
and antisymmetric $f_{A}(-{\sigma})=-f_{A}({\sigma})$ function
\begin{eqnarray}
f_{S}(\bar{\sigma})\delta^{\prime}_{A}(\sigma,\bar{\sigma})&=&
f_{S}({\sigma})\delta^{\prime}_{A}(\sigma,\bar{\sigma})
+f^{\prime}_{S}({\sigma})\delta_{A}(\sigma,\bar{\sigma}),
\nonumber\\
f_{A}(\bar{\sigma})\delta^{\prime}_{A}(\sigma,\bar{\sigma})&=&
f_{A}({\sigma})\delta^{\prime}_{S}(\sigma,\bar{\sigma})
+f^{\prime}_{A}({\sigma})\delta_{S}(\sigma,\bar{\sigma}),
\nonumber\\
f_{A}(\bar{\sigma})\delta^{\prime}_{S}(\sigma,\bar{\sigma})&=&
f_{A}({\sigma})\delta^{\prime}_{A}(\sigma,\bar{\sigma})
+f^{\prime}_{A}({\sigma})\delta_{A}(\sigma,\bar{\sigma}),
\nonumber\\
f_{S}(\bar{\sigma})\delta^{\prime}_{S}(\sigma,\bar{\sigma})&=&
f_{S}({\sigma})\delta^{\prime}_{S}(\sigma,\bar{\sigma})
+f^{\prime}_{S}({\sigma})\delta_{S}(\sigma,\bar{\sigma}).
\end{eqnarray}

For more details see App. A of Ref. \cite{DS}.
%%%%%%%%%%%%%%%%%%%%%%%%%%%%%%%%%%%%%%%%%%%%%%%%%%%%%%%%%%%%%%%%%%%%%%%%%%%%%%%%%%%%
\section{Background fields}
\cleq
In this Appendix we will generalize the expressions and the
terminology of the Ref. \cite{SW} for different descriptions
of the open string theory.
For more details see the last paragraph of the Introduction.
%%%%%%%%%%%%%%%%%%%%%%%%%%%%%%%%%%%%%%%%%%%%%%%%%%%%%%%%%%%%%%%%%%%%%%%%%%%%%%%%%%%
\subsection{Open string background fields}\label{sec:def}
Let us in analogy with the case of the flat background
introduce {\it open string background fields}
\begin{equation}\label{eq:gefv}
G^{E}_{\mu\nu}(G,B)\equiv G_{\mu\nu}-4B_{\mu\rho}(G^{-1})^{\rho\sigma}B_{\sigma\nu},
\end{equation}
\begin{equation}\label{eq:teta}
{\theta}^{\mu\nu}(G,B)
\equiv-\frac{2}{\kappa}\Big{[}G^{-1}_{E}(G,B)\Big{]}^{\mu\rho}
B_{\rho\sigma}(G^{-1})^{\sigma\nu}.
\end{equation}
When they depend on the
closed string parameters $G_{\mu\nu}$ and $B_{\mu\nu}$,
we will omit the arguments.

The corresponding constant parts
\begin{eqnarray}\label{eq:gtetao}
g_{\mu\nu}\equiv G_{\mu\nu}
-4b_{\mu\rho}(G^{-1})^{\rho\sigma}b_{\sigma\nu},
&&
\theta_{0}^{\mu\nu}\equiv-\frac{2}{\kappa}
(g^{-1})^{\mu\rho}b_{\rho\sigma}(G^{-1})^{\sigma\nu},
\end{eqnarray}
are related by expression
\begin{equation}\label{eq:pomoc}
\Big{(}G^{-1}b\theta_{0}\Big{)}^{\mu\nu}=
\frac{1}{2\kappa}\Big{(}G^{-1}-g^{-1}\Big{)}^{\mu\nu}.
\end{equation}
It is useful to introduce notation for infinitesimal parts of background fields
\begin{eqnarray}\label{eq:deltag}
\Delta G^{E}_{\mu\nu}\equiv G^{E}_{\mu\nu}- g_{\mu\nu},
&&
\Delta {\theta}^{\mu\nu}\equiv{\theta}^{\mu\nu}-{\theta}^{\mu\nu}_{0},
\end{eqnarray}
and for totally antisymmetric parameters
\begin{equation}\label{eq:tetatrig}
\theta^{\mu\nu\rho}\equiv\theta^{\mu\alpha}\partial_\alpha\theta^{\nu\rho}
+\theta^{\nu\alpha}\partial_\alpha\theta^{\rho\mu}
+\theta^{\rho\alpha}\partial_\alpha\theta^{\mu\nu},
\end{equation}
and
\begin{equation}\label{eq:tetatrid}
\theta_{\mu\nu\rho}\equiv
\partial_\mu\theta^{-1}_{\nu\rho}
+\partial_\nu\theta^{-1}_{\rho\mu}+\partial_\rho\theta^{-1}_{\mu\nu}.
\end{equation}
The $\theta_{\mu\nu\rho}$ and $\theta^{\mu\nu\rho}$ are constant and infinitesimally small,
because both $\theta^{\mu\nu}$ and $\theta^{-1}_{\mu\nu}$ are linear in $q$.
They are related by expression
\begin{equation}\label{eq:tgtd}
\theta^{\mu\nu\rho}=
\theta_{0}^{\mu\alpha}
\theta_{0}^{\nu\beta}
\theta_{0}^{\rho\gamma}
\theta_{\alpha\beta\gamma}.
\end{equation}
%%%%%%%%%%%%%%%%%%%%%%%%%%%%%%%%%%%%%%%%%%%%%%%%%%%%%%%%%%%%%%%%%%%%%%%%%%%
\subsection{Open string background fields
in terms of the effective fields}\label{sec:osbf}
Using (\ref{eq:gefv}) and (\ref{eq:teta}) we can define
new kind of
the open string backgrounds
\begin{eqnarray}\label{eq:gteff}
G^{E}_{\mu\nu}(G^{eff},B^{eff})&=&G^{eff}_{\mu\nu}-4B^{eff}_{\mu\rho}
(G_{eff}^{-1})^{\rho\sigma}B^{eff}_{\sigma\nu}
\equiv {G}^{E\ eff}_{\mu\nu},
\nonumber\\
{\theta}^{\mu\nu}(G^{eff},B^{eff})&=&
-\frac{2}{\kappa}\Big{[}G^{-1}_{E}(G^{eff},B^{eff})\Big{]}^{\mu\rho}
B^{eff}_{\rho\sigma}(G_{eff}^{-1})^{\sigma\nu}
\equiv {\theta}_{eff}^{\mu\nu},
\end{eqnarray}
in terms of the effective background fields
$G^{eff}$, $B^{eff}$
defined in (\ref{eq:gbtoeff}).

Taking into account that $B^{eff}$ is infinitesimal,
up to the first order we have
\begin{eqnarray}\label{eq:gtteff}
{{G}}^{E\ eff}_{\mu\nu}[q]=G^{eff}_{\mu\nu}[q],&&
{{\theta}}_{eff}^{\mu\nu}[{\bar{q}}_{0}]=-\frac{2}{\kappa}
[g^{-1}B_{eff}[\bar{q}_{0}]g^{-1}]^{\mu\nu}
=\Delta {{\theta}}^{\mu\nu}[{\bar{q}}_{0}],
\end{eqnarray}
where $\Delta\theta^{\mu\nu}$ is defined in (\ref{eq:deltag}).
Note that the new kind of open string metric
 ${G}^{E eff}_{\mu\nu}$  depends on effective coordinate $q^\mu$ while
the open string non-commutative parameter ${{\theta}}_{eff}^{\mu\nu}$
depends on the effective momenta,
or on
the T-dual effective coordinate
$\bar{q}_{0}^\mu=
-\theta_{0}^{\mu\nu}P_\nu
=2(G^{-1}b)^{\mu}_{\ \nu}{\tilde{q}}^\nu$.
%%%%%%%%%%%%%%%%%%%%%%%%%%%%%%%%%%%%%%%%%%%%%%%%%%%%%%%%%%%%%%%%%%%%%%%%%%%%%%%%%
\subsection{Particular combinations of the background fields}
In the description with the light-cone variables,
it is natural to define
\begin{equation}
\Pi_{\pm\mu\nu}\equiv B_{\mu\nu}\pm\frac{1}{2}G_{\mu\nu},
\end{equation}
and its analog of (\ref{eq:teta})
\begin{eqnarray}
{_{1}\Lambda}^{\mu\nu}_{\pm}[x]&\equiv&
-\frac{2}{\kappa}(G^{-1}_{E})^{\mu\alpha}\Pi_{\pm\alpha\beta}(G^{-1})^{\beta\nu}
\nonumber\\
&=&\theta^{\mu\nu}[x]\mp\frac{1}{\kappa}(G^{-1}_{E})^{\mu\nu}[x].
\end{eqnarray}

We also introduce similar linear combination where argument of $G^{-1}_{E}$
can be obtained by multiplying the argument of $\theta^{\mu\nu}$ by $b$
\begin{equation}\label{eq:lambda}
{{\Lambda}}^{\mu\nu}_{\pm}[q]\equiv\theta^{\mu\nu}[q]\mp
\frac{6}{\kappa}(G^{-1}_{E})^{\mu\nu}[bq],
\end{equation}
\begin{equation}\label{eq:barteta}
_{2}\Lambda^{\mu\nu}_{\pm}[q]\equiv
{\theta}^{\mu\nu}[q]
\mp\frac{2}{\kappa}(G^{-1}_{E})^{\mu\nu}[bq].
\end{equation}

All the above functions satisfy
\begin{equation}\label{eq:ttpm}
{{\Lambda}}^{\mu\nu}_{\pm}=-{\Lambda}^{\nu\mu}_{\mp}.
\end{equation}

%%%%%%%%%%%%%%%%%%%%%%%%%%%%%%%%%%%%%%%%%%%%%%%%%%%%%%%%%%%%%%%%%%%%%%%%%%%%%%%%%%%%%%%%%%%%%%
\subsection{First order expansion}
First let us expand the Kalb-Ramond field
in the first order in $B_{\mu\nu\rho}$
\begin{equation}\label{eq:KR}
B_{\mu\nu}[x]=
b_{\mu\nu}+h_{\mu\nu}[x],
\end{equation}
with the infinitesimal parts equal to
\begin{eqnarray}\label{eq:hhbar}
h_{\mu\nu}[x]\equiv\frac{1}{3}B_{\mu\nu\rho}x^\rho=h_{\mu\nu}+{\bar{h}}_{\mu\nu},&&
h_{\mu\nu}\equiv h_{\mu\nu}[q],\quad {\bar{h}}_{\mu\nu}\equiv h_{\mu\nu}[\bar{q}].
\end{eqnarray}
Consequently, the expressions defined in the
subsection \ref{sec:def} can be expanded as
\begin{eqnarray}\label{eq:gef}
G^{E}_{\mu\nu}[x]
&=&g_{\mu\nu}-4\Big{[}b(h+\bar{h})+(h+\bar{h})b\Big{]}_{\mu\nu},
\end{eqnarray}
\begin{eqnarray}\label{eq:tetared}
{\theta}^{\mu\nu}[x]
&=&{\theta}_{0}^{\mu\nu}
-\frac{2}{\kappa}\Big{[}g^{-1}
\Big{(}h+\bar{h}+4b(h+\bar{h})b\Big{)}
g^{-1}\Big{]}^{\mu\nu}
\nonumber\\
&=&
{\theta}_{0}^{\mu\nu}-\frac{2}{\kappa}[g^{-1}(h+\bar{h})g^{-1}]^{\mu\nu}
-2\kappa[\theta_{0}(h+\bar{h})\theta_{0}]^{\mu\nu},
\end{eqnarray}
and
\begin{eqnarray}\label{eq:gefin}
(G^{-1}_{E})^{\mu\nu}[x]&=&(g^{-1})^{\mu\nu}+4(g^{-1})^{\mu\rho}
\Big{[}b(h+\bar{h})+(h+\bar{h})b\Big{]}_{\rho\sigma}(g^{-1})^{\sigma\nu}
\nonumber\\
&=&(g^{-1})^{\mu\nu}
-2\kappa\Big{[}\theta_{0}(h+\bar{h})(g^{-1})
+g^{-1}(h+\bar{h})\theta_{0}\Big{]}^{\mu\nu},
\end{eqnarray}
\begin{eqnarray}\label{eq:tetaredin}
{\theta}^{-1}_{\mu\nu}[x]
&=&({\theta}_{0})^{-1}_{\mu\nu}
+\frac{\kappa}{2}\Big{\{}
Gb^{-1}\Big{[}
(h+\bar{h}+4b(h+\bar{h})b
\Big{]}b^{-1}G
\Big{\}}_{\mu\nu}.
\end{eqnarray}

%%%%%%%%%%%%%%%%%%%%%%%%%%%%%%%%%%%%%%%%%%%%%%%%
\section{Auxiliary functions}
To obtain the compact form for the infinite set of the constraints
we derive some useful expressions.
Relations of the subsection \ref{sec:kj} have been used in the
subsection \ref{sec:addterm}.
\subsection{Function $i^{\alpha\beta}$}\label{sec:funi}
\cleq
In this section we derive the expression for functions $i^{\alpha\beta}$ defined as
\begin{equation}\label{eq:idef}
i^{\alpha\beta}[a,b](\sigma)=
\frac{\sigma}{2}
\sum_{k=0}^{\infty}\frac{(-1)^{k}}{(k+1)!}
\int_{0}^{\sigma}d\sigma_{1}^{2}\cdots
\int_{0}^{\sigma_{k-1}}d\sigma_{k}^{2}
\int_{0}^{\sigma_{k}}d\eta
a^{(k)\alpha}(\eta) b^{(k+1)\beta}(\eta).
\end{equation}
The expression depends on $\sigma$-parity characteristics of the functions $a$ and $b$.
When the first variable is $\sigma$-symmetric and the second is $\sigma$-antisymmetric the function is equal to
\begin{equation}\label{eq:hsa}
i^{\alpha\beta}[a,\bar{b}](\sigma)=\frac{1}{2}{A}^\alpha(\sigma){\bar{b}}^{\beta}(\sigma),
\quad A^\alpha(\sigma)\equiv\int_{0}^{\sigma}d\eta a(\eta),
\end{equation}
and when the first variable is $\sigma$-antisymmetric and the second is $\sigma$-symmetric
\begin{equation}\label{eq:has}
i^{\alpha\beta}[\bar{a},b](\sigma)=0.
\end{equation}
To prove this statement we
substitute the expression for $\sigma$-even function
\begin{equation}
{a}^{(k)\alpha}(\eta)=2\int_{0}^{\pi}d\xi {a}^\alpha(\xi)\frac{\partial^{k}}{\partial\eta^{k}} \delta_{S}(\xi,\eta),
\end{equation}
into (\ref{eq:idef}) and use the following formula
\begin{eqnarray}
\int_{0}^{\sigma}d\eta f_{q}(\eta)\partial_{\eta}^{k}\delta_{S}(\rho,\eta)
=(-1)^{k}\partial^{k}_{\rho}[f_{q}(\rho)\theta_{S}(\sigma,\rho)],\qquad (k-q=2r),
\end{eqnarray}
for function $f_{q}$ with the property $f_{q}(-\eta)=(-1)^{q}f_{q}(\eta)$.
It produces
\begin{eqnarray}
i^{\alpha\beta}[a,\bar{b}](\sigma)=\sigma\int_{0}^{\pi}d\xi a^\alpha(\xi)Z^\beta(\bar{b}|\sigma,\xi),
\end{eqnarray}
and
\begin{eqnarray}
i^{\alpha\beta}[\bar{a},b](\sigma)=-\sigma\int_{0}^{\pi}d\xi b^\beta(\xi)\partial_\xi Z^\alpha(\bar{A}|\sigma,\xi),
\qquad \bar{A}(\sigma)\equiv\int_{0}^{\sigma}d\eta \bar{a}(\eta),
\end{eqnarray}
where $Z^\alpha(x|\sigma,\bar{\sigma})$ stands for
\begin{equation}\label{eq:zet}
Z^\alpha(x|\sigma,\bar{\sigma})\equiv\sum_{k=0}^{\infty}\frac{1}{(k+1)!}
\partial^{k}_{\bar{\sigma}}[x^{(k+1)\alpha}(\bar{\sigma})I_{k}(\sigma,\bar{\sigma})].
\end{equation}
Expressions (\ref{eq:hsa}) and (\ref{eq:has}) are obtained
after application of the result (\ref{eq:zexp}).

%%%%%%%%%%%%%%%%%%%%%%%%%%%%%%%%%%%%%%%%%%%%%%%%%%%%%%%%%%%%%%%%%%%%%%%%%%%%%%%%%%%%%%%%%%%%%%%%%%%%%

\subsubsection{Summation formula for $Z^\mu(x|\sigma,\bar{\sigma})$}\label{sec:zsum}
In this subsection we show that the $Z^\mu(x|\sigma,\bar{\sigma})$ defined in (\ref{eq:zet})
is equal to
\begin{equation}\label{eq:zexp}
Z^\mu(x|\sigma,\bar{\sigma})=\frac{1}{2\sigma}[x^\mu(\sigma)-x^\mu(-\sigma)]\theta_{S}(\sigma,\bar{\sigma}).
\end{equation}
In the paper \cite{DS} we proved the relation
\begin{eqnarray}\label{eq:sf}
S^\mu(x|\sigma,\bar{\sigma})&\equiv &\sum_{k=0}^{\infty}
\frac{1}{(k+1)!}
\partial^{k}_{\bar{\sigma}}
\Big{[}\bar{\sigma} x^{(k+1)\mu}(\bar{\sigma})I_{k}({\sigma},\bar{\sigma})\Big{]}
\nonumber\\
&=&\frac{1}{2}\theta_{S}(\sigma,\bar{\sigma})
[x^\mu(\sigma)+x^\mu(-\sigma)-2x^\mu(-\bar{\sigma})].
\end{eqnarray}
The expression (\ref{eq:zexp}) is obtained as a solution of the differential equation
\begin{equation}
\partial_{\sigma}S^\mu(x|\sigma,\bar{\sigma})=-\sigma
\Big{[}\partial_{\bar{\sigma}}Z^\mu(x|\sigma,\bar{\sigma})-Z^\mu(x^\prime|\sigma,\bar{\sigma})\Big{]}.
\end{equation}
%%%%%%%%%%%%%%%%%%%%%%%%%%%%%%%%%%%%%%%%%%%%%%%%%%%%%%%%%%%%%%%%%%%%%%%%%%%%%%%%%%%%%%%%%
\subsection{Integrals K and J}\label{sec:kj}
Integrating by parts it can be shown that
the integrals
\begin{eqnarray}
K_\rho(\sigma,\bar{\sigma})&=&
\int_{0}^{\sigma}d\eta
\int_{0}^{\bar{\sigma}}d\xi \delta_{S}(\eta,\xi)p_\rho(\eta),
\nonumber\\
J_\rho(\sigma,\bar{\sigma})&=&
\int_{0}^{\sigma}d\eta
\int_{0}^{\bar{\sigma}}d\xi \frac{\partial}{\partial\eta}
\delta_{S}(\eta,\xi)P_\rho(\eta),
\end{eqnarray}
obey the relations
\begin{eqnarray}\label{eq:ij}
K_\rho(\sigma,\bar{\sigma})+J_\rho(\sigma,\bar{\sigma})&=&
P_\rho(\sigma)\theta_{S}(\bar{\sigma},\sigma),
\nonumber\\
J_\rho(\sigma,\bar{\sigma})&=&
-P_\rho(\bar{\sigma})\theta_{S}(\sigma,\bar{\sigma}).
\end{eqnarray}

%%%%%%%%%%%%%%%%%%%%%%%%%%%%%%%%%%%%%%%%%%%%%%%%%%%%%%%%%%%%%%%%%%%%%%%%%%%%%%%%%%%%%%%%%%%%%%%%%%%%%%%%%%%%%%%%%%%

\end{document}